\documentclass[useAMS,usenatbib,usegraphicx]{mn2e}
\title[NGC~1817, NGC~2141, Berkeley~81]{NGC~1817, NGC~2141, and Berkeley~81: three BOCCE clusters of intermediate age\thanks{Based on observations collected at the Large Binocular Telescope
(LBT). The LBT is an international collaboration among institutions in the
United States, Italy and Germany. LBT Corporation partners are: The University
of Arizona on behalf of the Arizona University system; Istituto Nazionale di
Astrofisica, Italy; LBT Beteiligungsgesellschaft, Germany, representing the
Max-Planck Society, the Astrophysical Institute Potsdam, and Heidelberg
University; The Ohio State University, and The Research Corporation, on behalf
of The University of Notre Dame, University of Minnesota and University of
Virginia.}}
\author[Donati et al.]{P. Donati$^{1,2}$, G. Beccari$^{3}$, A. Bragaglia$^{2}$, M. Cignoni$^{1,2,4}$, M. 
Tosi$^{2}$
\\
$^{1}$Dipartimento di Fisica e Astronomia, via Ranzani 1, I-40127 Bologna, Italia\\
$^{2}$INAF-Osservatorio Astronomico di Bologna, via Ranzani 1, I-40127 Bologna, Italia\\
$^{3}$European Southern Observatory, Alonso de Cordova 3107, 19001 Santiago de Chile, Chile\\
$^{4}$Space Telescope Science Institute, Baltimore, MD, USA}

\begin{document}

\pagerange{\pageref{firstpage}--\pageref{lastpage}} 

\maketitle

\label{firstpage}

\begin{abstract}
In this paper we analyse the evolutionary status of three open clusters: NGC~1817, NGC~2141, and Berkeley~81. They are all of intermediate age, two are located in the Galactic anti-centre direction while the third one is located in the Galactic centre direction. All of them were observed with LBC@LBT using the Bessel B, V, and I filters. The cluster parameters have been obtained using the synthetic colour-magnitude diagram (CMD) method, i.e. the direct comparison of the observational CMDs with a library of synthetic CMDs generated with different evolutionary sets (Padova, FRANEC, and FST). This analysis shows that NGC~1817 has subsolar metallicity, age between 0.8 and 1.2 Gyr, reddening $E(B-V)$ in the range 0.21 and 0.34, and distance modulus $(m-M)_0$ of about 10.9; NGC~2141 is older, with age in the range 1.25 and 1.9 Gyr, $E(B-V)$ between 0.36 and 0.45, $(m-M)_0$ between 11.95 and 12.21, and subsolar metallicity; Berkeley~81 has metallicity about solar, with age between 0.75 and 1.0 Gyr, has reddening $E(B-V)\sim0.90$ and distance modulus $(m-M)_0\sim12.4$.
Exploiting the large field of view of the instrument we derive the structure parameters for NGC~2141 and Be~81 by fitting a King profile to the estimated density profile. Combining this information with the synthetic CMD technique we estimate a lower limit for the cluster total mass for these two systems.
\end{abstract}

\begin{keywords}
Hertzsprung-Russel and colour-magnitude diagrams, Galaxy: disc, open clusters and associations: general, open clusters and associations: individual: NGC 1817, open clusters and associations: individual: NGC 2141, open clusters and associations: individual: Berkeley 81.
\end{keywords}

\section{Introduction}
\label{sec:intro}
This paper is part of the long-term BOCCE (Bologna Open Clusters Chemical Evolution) project, 
aimed at precisely and homogeneously derive the fundamental properties of a large sample of Open Clusters (OCs), and described in detail by \cite{boc_06}. The ultimate goal  of the BOCCE project is  to get insight on the formation and evolution of the Galactic disc, and OCs are among the best tracers of the disc properties (e.g. \citealt{fri_95}). 
We have already analysed photometric data for 31 OCs (see \citealt{boc_06,cig_11,donati12,ahumada13},  and references therein), and derived their age, distance and reddening from the comparison of their colour-magnitude diagrams (CMDs) with synthetic ones based on three sets of stellar evolution models (see \citealt{boc_06}).

In this paper we discuss NGC~1817 (Galactic coordinates $l=207.8^{\circ},b=2.6^{\circ}$), NGC~2141 ($l=214.2^{\circ},b=1.9^{\circ}$), and Berkeley 81 (Be~81, $l=227.5^{\circ},b=-0.6^{\circ}$). 
These clusters have been selected because they could be targets of the Gaia-ESO Survey (GES, see \citealt{ges} for a description).

The GES is an on-going public spectroscopic survey with FLAMES@VLT, that will obtain high-resolution GIRAFFE and UVES spectra of about 10$^5$ stars of all Milky Way components, 
including stars in about 100 OCs and associations. For all the GES cluster targets we need photometry and precise astrometry covering all the FLAMES field of view (diameter of 25\arcmin) to properly point the fibres. Such adequate photometry  was not yet available for  NGC~2141, Be~81, and NGC~1817, and we acquired it on purpose with LBT.

NGC~2141 and NGC~1817 are anti-centre clusters, whilst Be~81 lies towards the Galactic centre\footnote{On the basis of the moduli derived in the following Sections their distances from the Galactic centre are  $R_{GC}\simeq9.5$ kpc for NGC~1817, $R_{GC}\simeq12$ kpc for NGC~2141, and $R_{GC}\simeq5.7$ kpc for Be~81. }, so they are particularly interesting to study the radial distribution of the disc properties. In Table~\ref{tablit} we report a consistent summary of all the parameters available in the literature for the three clusters. It is apparent that 
they  do not agree with each other, and a more precise analysis is called for.

{\em NGC~1817 -} Its richness, 
distance from the Galactic plane (-400 pc), and metallicity make this cluster particularly interesting. In fact, NGC~1817 has been the target of many photometric studies,
starting from \cite{ac62} and \cite{purgathofer61}, who obtained shallow
photographic CMDs, including only stars at the main sequence turn-off (MSTO) and some giants. \cite{harris77} acquired photographic $UBV$ data, providing a well defined MS and red clump (RC), and derived distance, reddening (see Table~\ref{tablit}),  age similar to the Hyades, and a low
metallicity. \cite{balaguer04a} performed deep, wide field photometry in the Str\"omgren system ($uvby-H\beta$), covering an area of $65\times40$ arcmin$^2$
and building on the proper motion and membership analysis by \cite{balaguer98}.
For the cluster members they derived the parameters listed in Table~\ref{tablit}.  A subsolar metallicity was derived by \cite{parisi05} on the basis of
Washington photometry. 

Spectroscopic analyses were made using low resolution spectra by \cite{friel93} and high resolution ones by \cite{jacobson09} and \cite{jacobson11} for different cluster stars. Despite showing different results all these studies point to a slightly subsolar metallicity (see Table~\ref{tablit}).
Crucial information on radial velocities (RVs), membership, and binary stars were given by
\citet{mermilliod03,mermilliod07}, and \cite{mermilliod08}.

{\em NGC~2141 -} It is a rich cluster, subject of several studies in the past.
\cite{burkhead72} obtained photoelectric and photographic $UBV$ data, barely
reaching below the MSTO; 
they determined the distance modulus and reddening listed in Table~\ref{tablit}, and an age intermediate between those of
M67 and NGC~2477.  \cite{rosvick95} observed an area of 173 arcmin$^2$
with $VI$ filters and a
smaller area with $JHK$. Her CMD reached about four
magnitudes below the MSTO, and showed a large scatter, interpreted in terms of  both field star
contamination  and
differential reddening. 
\cite{rosvick95}  determined the reddening,
distance modulus, metallicity and age listed in Table~\ref{tablit}
from a fit with the \cite{bertelli94} isochrones.  The latest photometric data for this cluster have been presented by \cite{carraro01}, who acquired
$BV$ and $JK$ data. Their optical CMD extends to
$V\sim21.5$, while the IR CMD reaches about two magnitudes below the MSTO.
They estimated the metallicity from the IR photometry, deriving best-fit age and distance, based on the
\cite{girardi00} isochrones (see Table~\ref{tablit}).

Spectroscopic analyses of cluster stars was made by different authors: \cite{friel93}, \cite{minniti95} used low resolution spectra while \cite{yong05} and \cite{jacobson09} high resolution ones. They found different values for the cluster metallicity from solar to sub-solar (see Table~\ref{tablit}). \cite{jacobson09} discussed the possible sources for the
discrepancy and thoroughly analysed 
the literature findings. 
In summary, this cluster has a metallicity near solar or slightly
lower, and this information  will be used here to constrain the
choice of the cluster's parameter.

{\em Berkeley~81 -} $BVI$ photometry of part of Be~81 has been presented by
\cite{sagar98}. They argued for the absence of
significant differential reddening from the CMDs of different regions, and
attributed the width of the MS to the presence of field stars, binaries, and
variables. They derived a cluster radius of $2.7\pm0.2$ arcmin, and the reddening, distance modulus and age listed in Table~\ref{tablit}, 
using the \cite{bertelli94} isochrones with solar metallicity.

The metallicity of  Be~81 was determined from  calcium
triplet (CaT) spectroscopy by \cite{wc09}. Their subsolar value is however quite uncertain, since
they were unable to convincingly define the cluster mean RV, due to the
huge contamination by field stars. The GES spectra will thus be crucial to infer its actual metallicity.

\begin{table*}
\label{tablit}
\caption{List of the main properties of the three clusters found in literature. The true distance modulus $(m-M)_0$ is evaluated from literature values after applying the same extinction law adopted in this paper ($R_V=3.1$).}
\begin{tabular}{lccccl}
\hline
  Cluster    & E(B-V)        &$(m-M)_0$     &     age   &metallicity   &Reference \\
\hline
NGC 1817    & 0.28       &11.3$\pm0.4$   &$\sim$Hyades &less than Hyades     &Harris \& Harris (1977) \\
             & 0.27       &10.9$\pm0.6$   & 1.1 Gyr &[Fe/H]$=-0.34\pm0.26$  &Balaguer-Nun\'ez et al. (2004) \\
         &           &           & &[Fe/H]$=-0.33\pm0.09$  &Parisi et al. (2005)\\
         &           &           & &[Fe/H]$=-0.38\pm0.04$  &Friel \& Janes (1993)\\
         &           &           & &[Fe/H]$=-0.07\pm0.04$  &Jacobson et al. (2009)\\
             &           &           & &[Fe/H]$=-0.16\pm0.03$  &Jacobson et al. (2011)\\
NGC 2141    & 0.3       &13.17 &NGC2477$<$age$<$M67 &         &Burkhead et al. (1972)\\
             &0.35$\pm0.07$ &13.08$\pm0.16$ &2.5 Gyr   &Z=0.004-0.008  &Rosvick (1995)\\
         &0.40       &12.90$\pm0.15$ &2.5 Gyr &[Fe/H]$=-0.43\pm0.07$  &Carraro et al. (2001)\\
         &           &           & &[Fe/H]$=-0.39\pm0.11$  &Friel \& Janes (1993)\\
         &           &           & &[Fe/H]$=-0.18\pm0.15$  &Yong et al. (2005)\\
         &           &           & &[Fe/H]$=+0.00\pm0.16$     &Jacobson et al. (2009)\\
Berkeley 81 &1.0           &12.5           &1 Gyr  &solar                  &Sagara \& Griffiths (1998)\\
             &           &           & &[Fe/H]$=-0.15\pm0.11$: &Warren \& Cole (2009)\\
\hline
\end{tabular}
\end{table*}

This paper is organised as follows. Observations and the resulting CMDs are
presented in Sec.~\ref{sec:data}; the estimation of the clusters centre in
Sec.~\ref{sec:centre}; differential reddening is discussed in Sec.~\ref{sec:diffredd}; the derivation of their age, distance, reddening, and
metallicity using comparison to synthetic CMDs in Sec.~\ref{sec:CMDsynth}.
Discussion and summary can be found in Sec.~\ref{sec:sum}.

\section[]{The Data}
\label{sec:data}
\begin{figure*}
\centering
\includegraphics[width=0.3\textwidth]{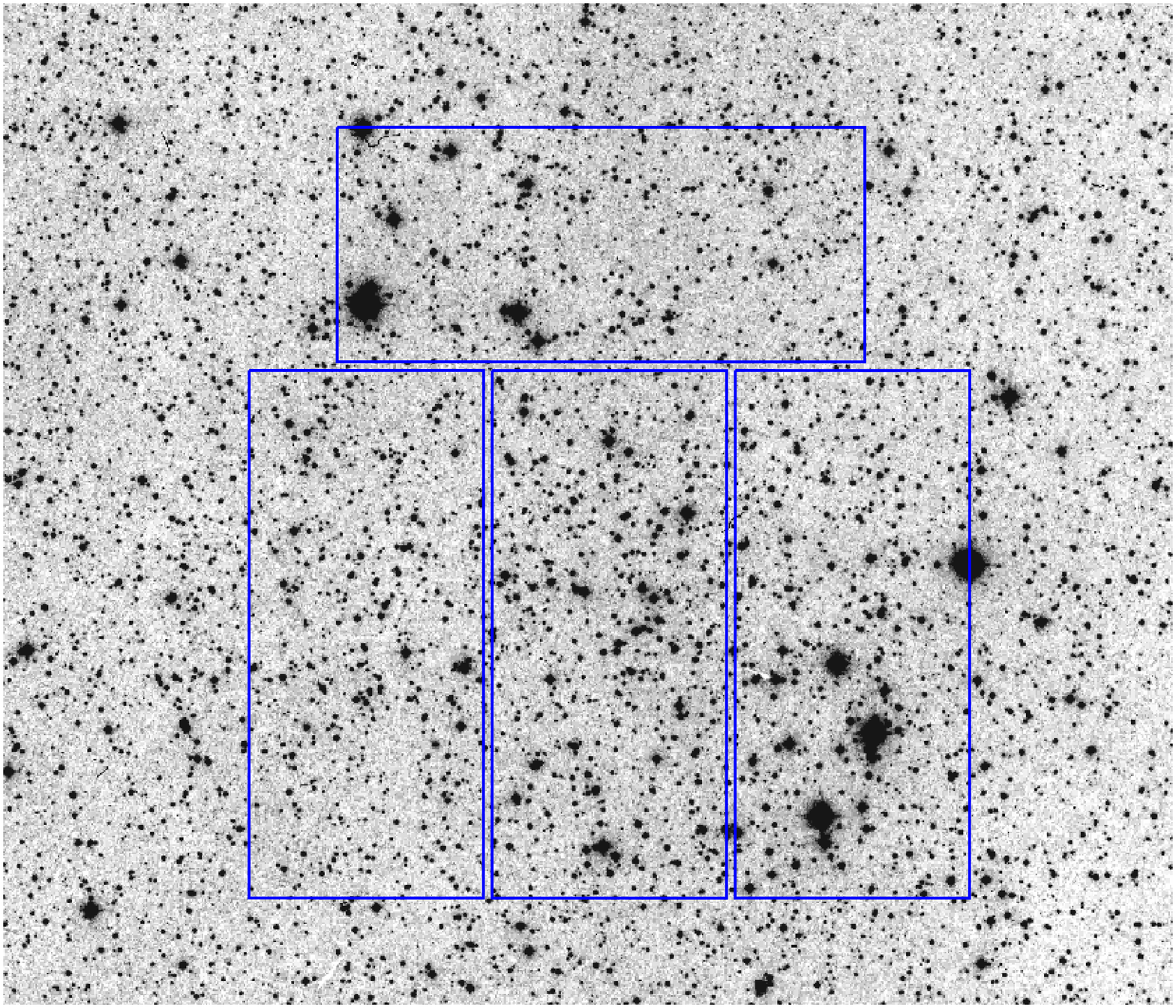}
\includegraphics[width=0.3\textwidth]{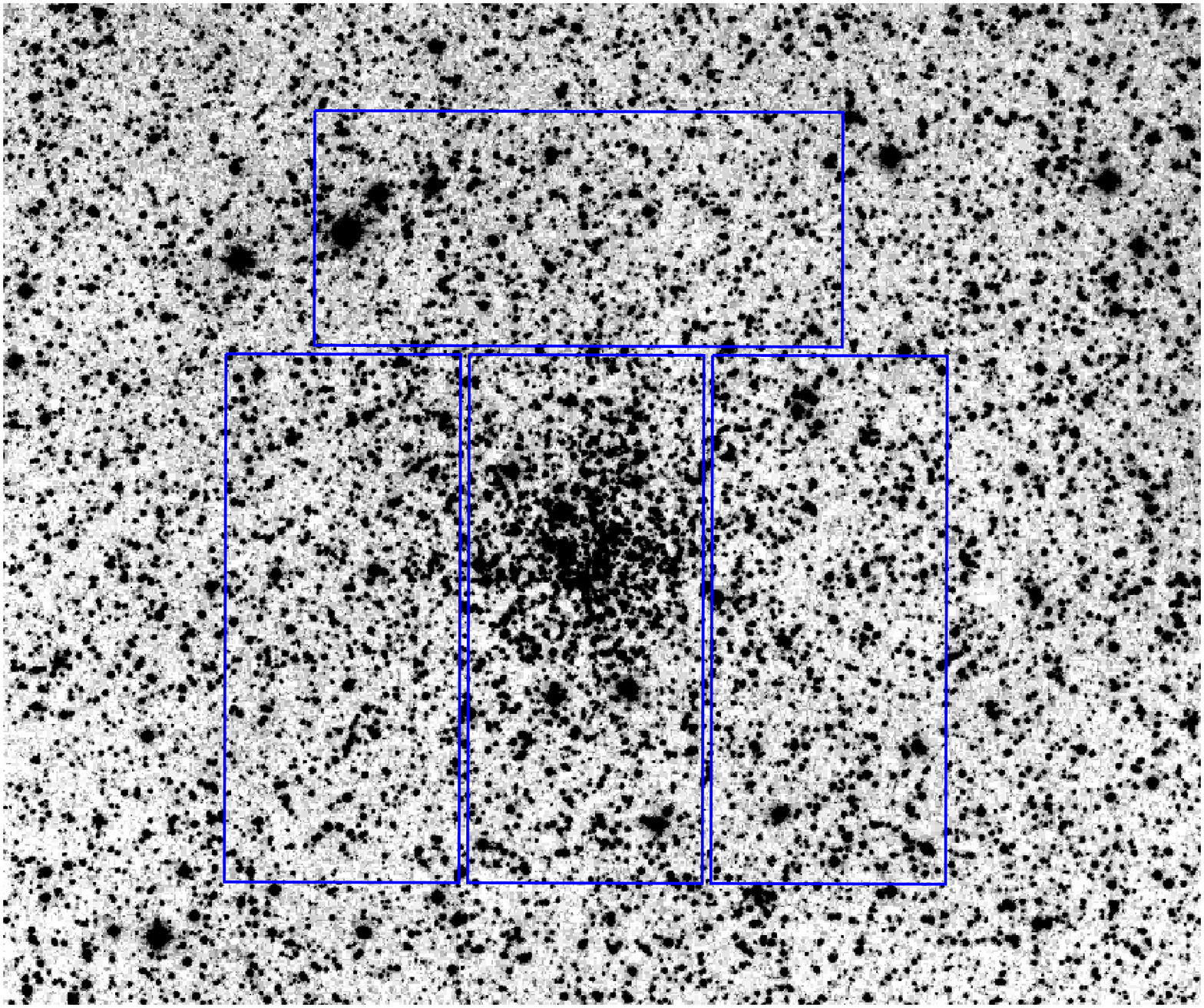}
\includegraphics[width=0.3\textwidth]{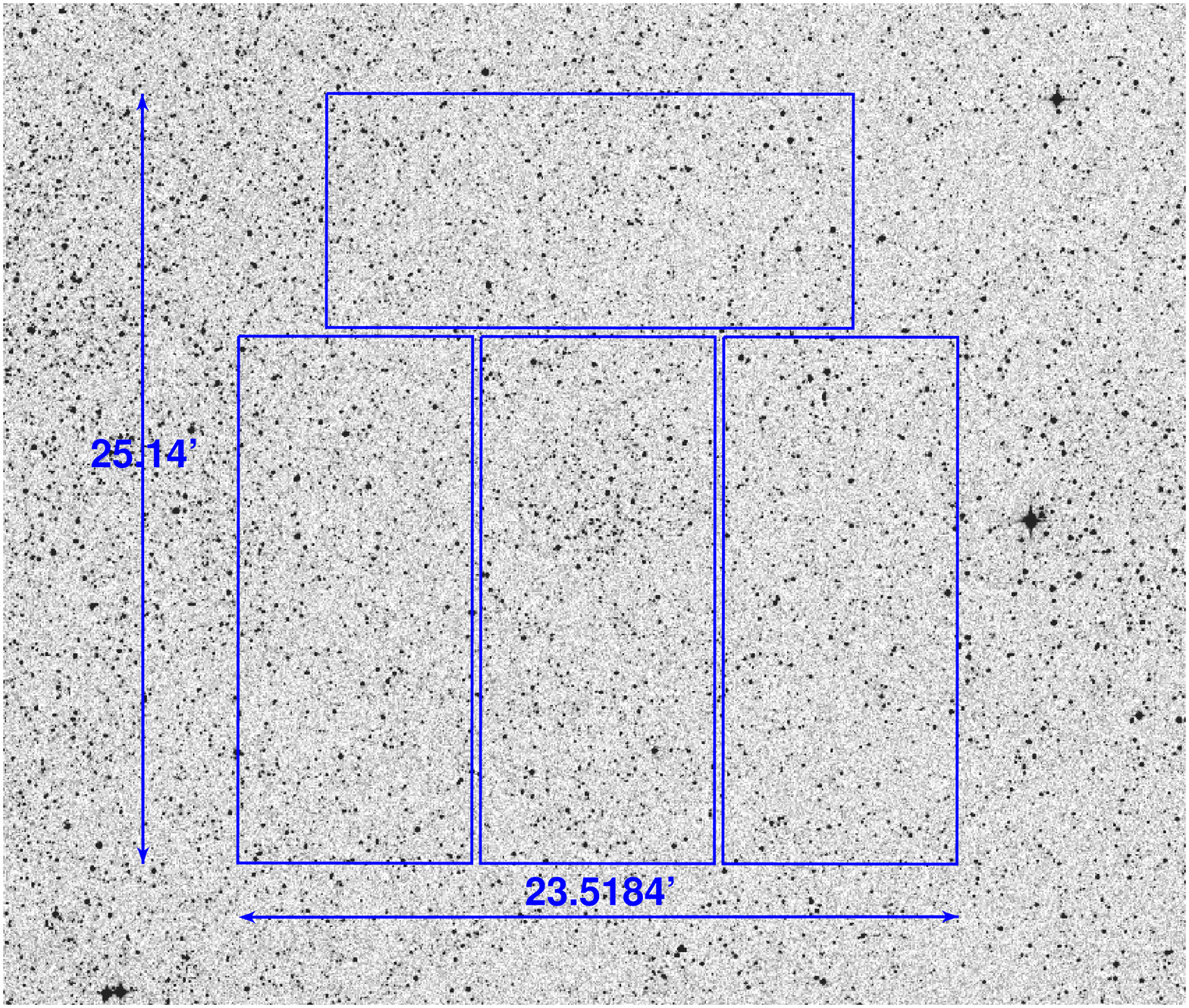}
\caption{The field of views of the three clusters, NGC~1817, NGC~2141, and Be~81, from left to right. In the last map on the right we highlight the dimension of the FoV in arcminutes. All these images were downloaded from the DSS SAO catalogue in the GSSS bandpass 6 (V495). North is up and East is left.}
\label{fig:map}
\end{figure*}

The three clusters were observed in service mode at the 
LBT
 on Mt. Graham (Arizona) with the Large Binocular Camera (LBC) in 2011 (see Tab. \ref{tab:logbook} for details). There are two LBCs, one optimised for the UV-blue filters and one for the red-IR ones, mounted at each prime focus of the LBT. Each LBC uses four EEV chips (2048$\times$4608 pixels) placed three in a row, and the fourth 
  above them and rotated by 90 deg (see Figure \ref{fig:map}).
  The field of view (FoV) of LBC is equivalent to $22\arcmin\times25\arcmin$, with a pixel sampling of 0.23$\arcsec$. The clusters were positioned in the central chip (\# 2) of the LBCs CCD mosaic (see Fig. \ref{fig:map}). We 
   observed in the $B$ filter with the LBC-Blue camera and in
   $V$ and $I$ with the LBC-Red one. No dithering pattern was adopted. Tab. \ref{tab:logbook} gives the log of the observations. The seeing was good (about 1$\arcsec$), and the airmass of the exposures was in the range 1.0-1.3. Landolt fields were observed to perform our own calibration to the Johnson-Cousins system.

\begin{table*}
 \centering
\caption{Logbook of the observations. The listed coordinates refer to the telescope pointings.}
\begin{tabular}{lcclcccl}
  \hline
   Cluster Name   &   RA (h m s)  & Dec ($\degr$ $\arcmin$  $\arcsec$) & Date & B & V & I & seeing \\
        & J2000 & J2000 &  & Exp. time & Exp. time &   Exp. time  & $\arcsec$   \\
 \hline
NGC~1817	& 05 12 41  &  16 44 30 & 24 Oct 2011 & 1s, 3$\times$5s, 3$\times$90s & 1s, 3$\times$5s, 3$\times$60s & 1s, 3$\times$5s, 3$\times$60s & 1$\arcsec$ \\
NGC~2141	& 06 02 57  &  10 27 27 & 21 Oct 2011 & 1s, 3$\times$5s, 3$\times$90s & 1s, 3$\times$5s, 3$\times$60s & 1s, 3$\times$5s, 3$\times$60s & 1$\arcsec$ \\
Be~81	& 19 01 41 & -00 27 40  & 20 Oct 2011 & 1s, 3$\times$5s, 3$\times$90s & 1s, 3$\times$5s, 3$\times$60s & 1s, 3$\times$5s, 3$\times$60s & 1$\arcsec$ \\
\hline
\end{tabular}
\label{tab:logbook}
\end{table*}

\begin{table*}
  \centering \caption{Completeness level for calibrated $B$, $V$, and $I$ magnitudes.}
  \begin{tabular}{c|c|c|c|c|c|c|c|c|c|}
    \hline \hline
    & \multicolumn{3}{c}{NGC~1817 (d$<5\arcmin$)} & \multicolumn{3}{c}{NGC~2141 (d$<4\arcmin$)} & \multicolumn{3}{c}{Be~81 ($d<2\arcmin$)}\\
    \hline
    bin & $B$ & $V$ & $I$ & $B$ & $V$ & $I$ & $B$ & $V$ & $I$ \\
    16.5 & -                 & 100.0 $\pm$   1.0 & -                 & -                 & 100.0 $\pm$   1.3 & 100.0 $\pm$   -   & -                 & -                 & 100.0 $\pm$   3.0  \\
    17.0 & 100.0 $\pm$   1.7 & 100.0 $\pm$   1.8 & 100.0 $\pm$   1.2 & -                 &  97.1 $\pm$   1.4 & 100.0 $\pm$   1.2 & -                 & 100.0 $\pm$   6.0 &  84.9 $\pm$   2.9  \\
    17.5 & 100.0 $\pm$   1.4 & 100.0 $\pm$   1.5 &  95.3 $\pm$   1.1 & 100.0 $\pm$   1.2 &  97.1 $\pm$   1.4 &  95.7 $\pm$   1.1 & -                 &  97.7 $\pm$   5.5 &  77.6 $\pm$   2.2  \\
    18.0 & 100.0 $\pm$   1.9 &  96.1 $\pm$   1.4 &  94.4 $\pm$   1.0 &  97.4 $\pm$   1.4 &  97.4 $\pm$   1.2 &  94.7 $\pm$   1.0 & 100.0 $\pm$   8.2 &  94.2 $\pm$   3.9 &  78.1 $\pm$   1.9  \\
    18.5 & 100.0 $\pm$   1.6 &  95.4 $\pm$   1.4 &  92.5 $\pm$   1.0 &  97.1 $\pm$   1.4 &  96.7 $\pm$   1.2 &  93.7 $\pm$   0.9 &  98.5 $\pm$   7.5 &  93.4 $\pm$   3.6 &  69.6 $\pm$   1.4  \\
    19.0 &  96.0 $\pm$   1.6 &  95.9 $\pm$   1.3 &  87.5 $\pm$   0.9 &  96.8 $\pm$   1.2 &  95.9 $\pm$   1.2 &  90.8 $\pm$   0.9 &  95.2 $\pm$   4.5 &  92.2 $\pm$   2.9 &  64.9 $\pm$   1.1  \\
    19.5 &  95.5 $\pm$   1.5 &  95.0 $\pm$   1.3 &  83.7 $\pm$   0.8 &  96.6 $\pm$   1.3 &  94.6 $\pm$   1.1 &  81.1 $\pm$   0.8 &  93.4 $\pm$   3.9 &  89.9 $\pm$   2.5 &  54.6 $\pm$   0.9  \\
    20.0 &  95.5 $\pm$   1.5 &  94.4 $\pm$   1.2 &  68.9 $\pm$   0.8 &  96.0 $\pm$   1.2 &  93.7 $\pm$   1.1 &  69.4 $\pm$   0.7 &  93.2 $\pm$   3.5 &  88.4 $\pm$   2.1 &  35.0 $\pm$   0.6  \\
    20.5 &  95.5 $\pm$   1.4 &  92.4 $\pm$   1.2 &  28.5 $\pm$   0.4 &  94.6 $\pm$   1.2 &  91.4 $\pm$   1.0 &  28.0 $\pm$   0.4 &  92.0 $\pm$   3.0 &  82.4 $\pm$   1.6 &   9.7 $\pm$   0.3  \\
    21.0 &  95.1 $\pm$   1.4 &  88.8 $\pm$   1.1 &   3.0 $\pm$   0.1 &  94.7 $\pm$   1.2 &  83.5 $\pm$   1.0 &   1.6 $\pm$   0.1 &  88.6 $\pm$   2.5 &  77.7 $\pm$   1.3 &   1.1 $\pm$   0.1  \\
    21.5 &  94.7 $\pm$   1.4 &  84.0 $\pm$   1.0 &   0.2 $\pm$   0.0 &  92.0 $\pm$   1.1 &  74.7 $\pm$   0.9 &   0.1 $\pm$   0.0 &  88.1 $\pm$   2.1 &  69.4 $\pm$   1.1 &   0.2 $\pm$   0.0  \\
    22.0 &  92.3 $\pm$   1.3 &  78.5 $\pm$   0.9 &   0.0 $\pm$   -   &  88.3 $\pm$   1.1 &  60.5 $\pm$   0.8 & -                 &  81.4 $\pm$   1.7 &  51.4 $\pm$   0.8 & -                  \\
    22.5 &  89.8 $\pm$   1.2 &  64.8 $\pm$   0.8 &   -               &  80.5 $\pm$   1.0 &  25.5 $\pm$   0.4 &  -                &  78.0 $\pm$   1.4 &  21.1 $\pm$   0.5 &  -                 \\
    23.0 &  85.2 $\pm$   1.0 &  31.2 $\pm$   0.5 &  -                &  74.9 $\pm$   1.0 &   2.6 $\pm$   0.1 &  -                &  71.6 $\pm$   1.2 &   3.1 $\pm$   0.2 &  -                 \\
    23.5 &  81.5 $\pm$   1.0 &   4.6 $\pm$   0.2 &  -                &  64.5 $\pm$   0.9 &   0.1 $\pm$   0.0 &  -                &  58.4 $\pm$   1.0 &   0.2 $\pm$   0.0 &  -                 \\
    24.0 &  72.5 $\pm$   0.9 &   0.3 $\pm$   -   &  -                &  40.7 $\pm$   0.7 &  -                &  -                &  36.2 $\pm$   0.7 & -                 &  -                 \\
    24.5 &  47.4 $\pm$   0.6 &   -               &  -                &   6.9 $\pm$   0.2 &  -                &  -                &   8.2 $\pm$   0.3 &  -                &  -                 \\
    25.0 &   8.0 $\pm$   0.2 &   -               &  -                &   0.3 $\pm$   0.0 &  -                &  -                &   0.3 $\pm$   0.1 &  -                &  -                 \\
    25.5 &   0.3 $\pm$   0.0 &  -                &  -                &  -                &  -                &  -                &  -                &  -                &  -                 \\
   \hline
    \hline
  \end{tabular}\label{tab:compl}
\end{table*}

\subsection{Data reduction}
The raw LBC images were corrected for bias and flat field, and the overscan region was trimmed using a pipeline specifically developed for LBC image prereduction by the Large Survey Center (LSC) team at the Rome Astronomical Observatory\footnote{LSC website: http://lsc.oa-roma.inaf.it/}. The source detection and relative photometry was performed independently on each B, V, and I image, using the PSF-fitting code DAOPHOTII/ALLSTAR \citep[][]{ste_87,ste_94}. We sampled the PSF using the highest degree of spatial variability allowed by the programme because the images are affected by severe spatial distortion. This procedure is adopted in other papers of this series and is proven to be effective to well sample the PSF on the whole frame. \cite{gial08} showed that the geometric distortion, of pin-cushion type, is always below 1.75\% even at the edge of the field. At any rate, for our purposes we mostly use the inner area of the FoV where a distortion up to only 1\% is expected. Moreover, the energy concentration of the instrumental PSF is very good: 80\% of the energy is enclosed in a single CCD pixel in the $B$ band and in 2$\times$2 pixels in the $V,I$ bands. 

The brightest stars, saturated in the deepest images, where efficiently recovered from the short exposure images.  The weighted average of the independent measures obtained from the different images were adopted as the final values of the instrumental magnitude (basing the weight on the error).
More than 200 stars from the 2MASS catalogue \citep{2mass} where used as astrometric standards to find an accurate astrometric solution 
and transform the instrumental positions, in pixels, into J2000 celestial coordinates for each chip. To this aim we adopted the code CataXcorr, developed by Paolo Montegriffo at 
the INAF - Osservatorio Astronomico di Bologna.
The rms scatter of the solution was about 0.1$\arcsec$ in both RA and Dec.

We derived the completeness level of the photometry by means of extensive artificial stars experiments following the recipe described in \cite{bel_02} and adopted in other papers of this series. About $10^5$ artificial stars were used to derive photometric errors and completeness in $B$, $V$, and $I$ exposures for the central chip. The results are shown in Tab. \ref{tab:compl}.

\subsection{Calibration and comparison with previous data}\label{sec:calib}
The calibration to the Johnson-Cousins photometric system was obtained using standard stars \citep{lan_92} obtained in the same observing nights. Landolt fields SA98, SA101, SA113L1, and L92 were observed at different airmasses: in the range 1.2-1.9 during the nights of Oct. 20 2011 and Oct. 21 2011, and in the range 1.2-1.5 during the third night.  It was not possible to derive a calibration equation for each chip. So, we used the same one for all the four CCDs.
The adopted calibration equation is the following:
$$(M-m_i)=zp+k\times C_i$$
where $M$ is the magnitude in the standard photometric system, $m_i$ the instrumental magnitude, $zp$ the zero point, and $k$ describes the linear dependence from the instrumental colour $C_i$. 
We adopted the average coefficients $k_B=-0.22$, $k_V=-0.15$, and $k_I=-0.04$ given by  the telescope web page for all the three clusters.
The results are summarised in Tab. \ref{tab:calib}.

\begin{table*}
 \centering
\caption{Calibration equations obtained for the three observing nights. The quoted zero-points include the zero-point adopted by DAOPHOTII (25 mag).}
\begin{tabular}{ccc}
\hline
\hline
\multicolumn{3}{c}{Oct. 20 2011} \\
\multicolumn{3}{c}{\# of individual CCD images: 184 in $B$, 188 in $V$, and 64 in $I$.}\\
\hline
equation & rms & stars used for each chip (1 to 4)\\
\hline
$B-b=2.696-0.111\times(b-v)$ & rms 0.01 & 148, 124, 64, and 164\\
$V-v=2.558-0.025\times(b-v)$ & rms 0.01 & 148, 124, 64, and 164\\
$V-v=2.558-0.032\times(v-i)$ & rms 0.02 & 93, 118, 86, and 100\\
$I-i=2.317+0.016\times(v-i)$ & rms 0.02 & 93, 118, 86, and 100\\
\hline
\multicolumn{3}{c}{Oct. 21 2011} \\
\multicolumn{3}{c}{\# of individual CCD images: 72 in $B$, 72 in $V$, and 40 in $I$}\\
\hline
equation & rms & stars used for each chip (1 to 4)\\
\hline
$B-b=2.715-0.107\times(b-v)$ & rms 0.03 & 61, 99, 78, and 53\\
$V-v=2.614-0.045\times(b-v)$ & rms 0.03 & 61, 99, 78, and 53\\
$V-v=2.615-0.053\times(v-i)$ & rms 0.03 & 104, 74, 107, and 70\\
$I-i=2.368-0.014\times(v-i)$ & rms 0.03 & 104, 74, 107, and 70\\
\hline
\multicolumn{3}{c}{Oct. 24 2011} \\
\multicolumn{3}{c}{\# of individual CCD images: 59 in $B$, 95 in $V$, and 63 in $I$}\\
\hline
equation & rms & stars used for each chip (1 to 4)\\
\hline
$B-b=2.759-0.209\times(b-v)$ & rms 0.02 & 11, 21, 11, and 9\\
$V-v=2.606-0.075\times(b-v)$ & rms 0.02 & 11, 21, 11, and 9\\
$V-v=2.576-0.039\times(v-i)$ & rms 0.03 & 29, 24, 14, and 16\\
$I-i=2.324+0.012\times(v-i)$ & rms 0.03 & 29, 24, 14, and 16\\
\hline
\end{tabular}
\label{tab:calib}
\end{table*}

\begin{figure}
\includegraphics[scale=0.45]{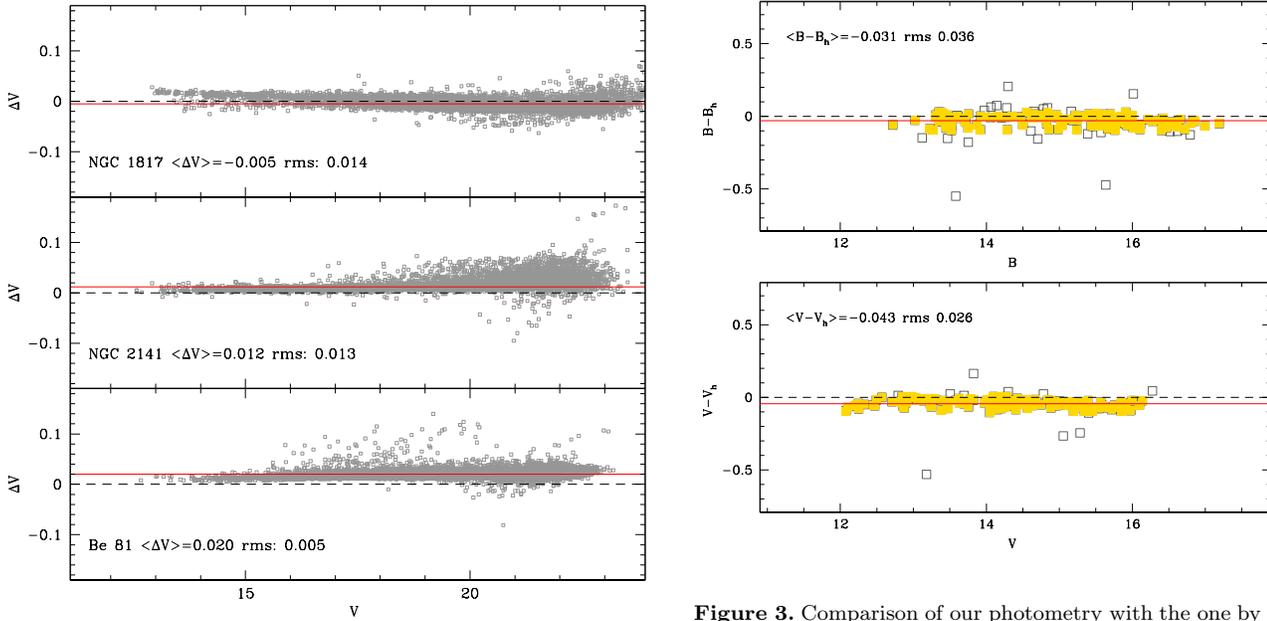}
\caption{Comparison of the $V$ calibrated from $(b-v)$ and $(v-i)$ colours with respect to the $V$ magnitude for the three clusters (from top to bottom NGC~1817, NGC~2141, and Be~81). The labelled values $<\Delta V>$ are the medians of all the stars shown for each plot.}
\label{fig:vsv}
\end{figure}

Comparing the calibrated $V$ obtained from 
$(b-v)$ with that  obtained with $(v-i)$, we find a small difference of $\leq0.02$ mag, which tends to worsen towards fainter magnitudes (see Fig. \ref{fig:vsv}). 
 
\begin{figure}
\includegraphics[scale=0.4]{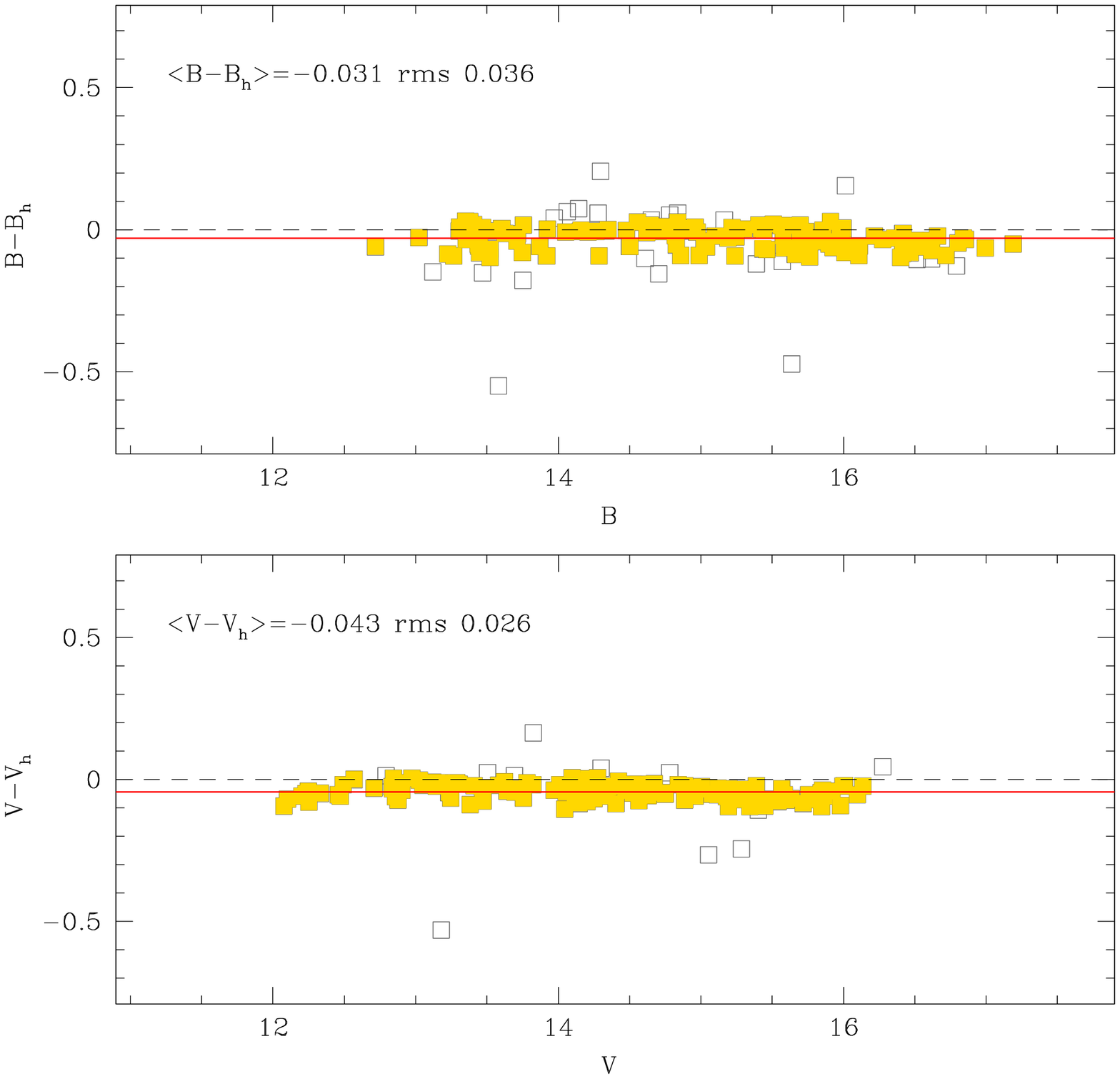}
\caption{Comparison of our photometry with the one by Harris et al. (1977) for NGC~1817. The average difference is computed using the golden points, retained after one sigma-clipping has been applied.}
\label{fig:vsLn1817}
\end{figure}

\begin{figure}
\includegraphics[scale=0.4]{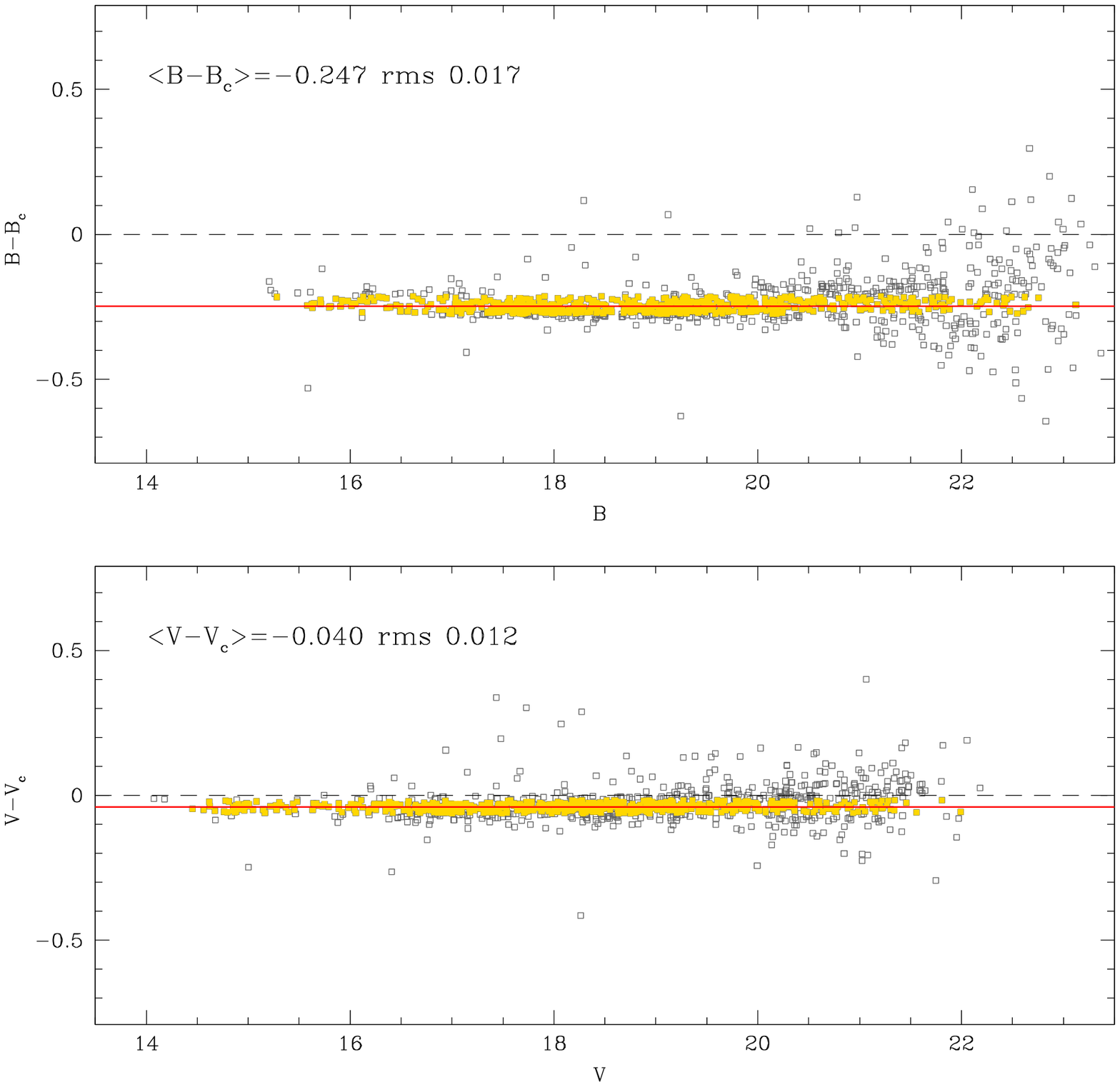}
\includegraphics[scale=0.4]{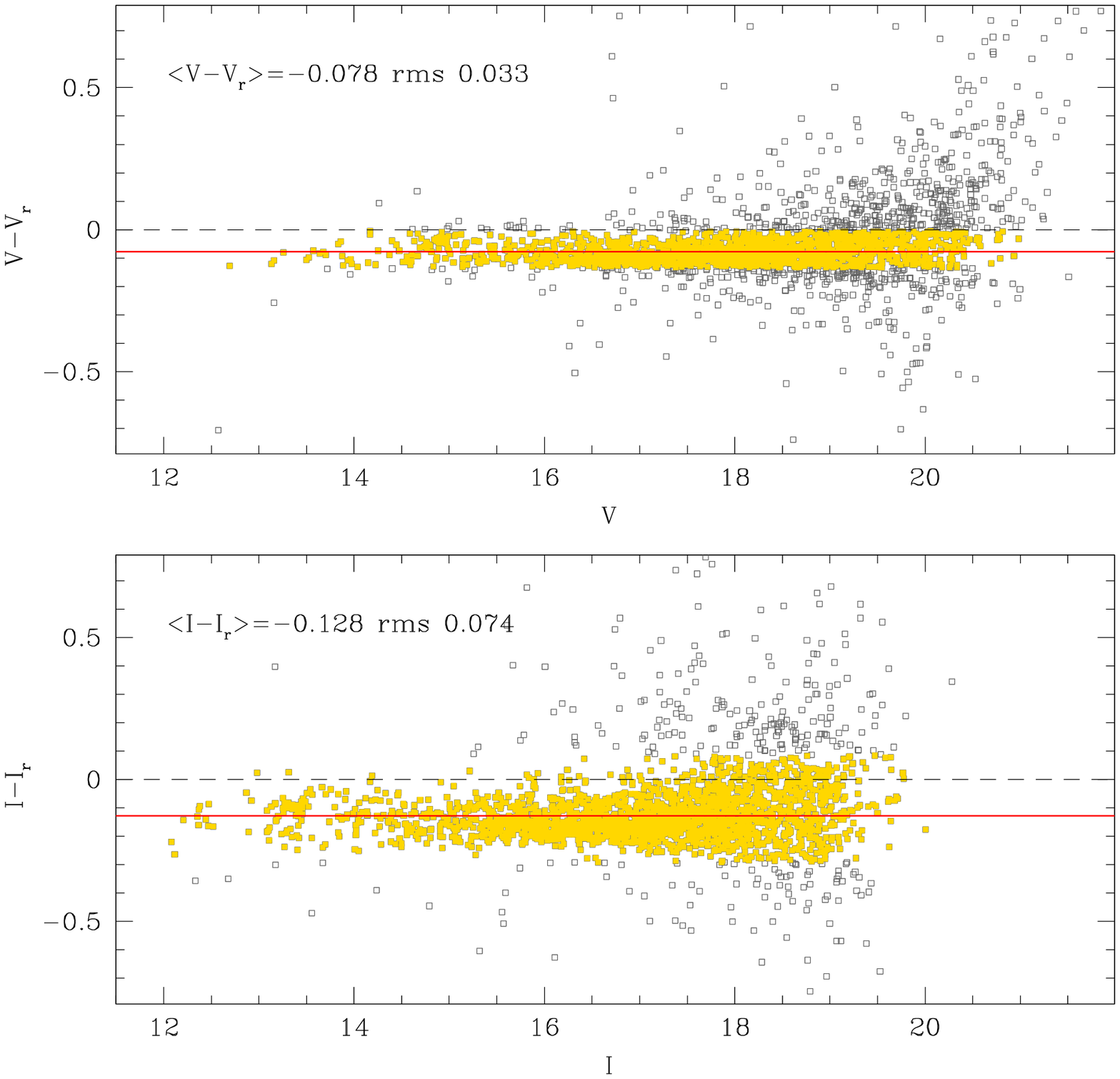}
\caption{Same as Fig. \ref{fig:vsLn1817}, but for NGC~2141. In the upper panels we compare $B$ and $V$ with the photometry of Carraro et al. (2001), in the bottom panels $V$ and $I$ with Rosvick et al. (1995).}
\label{fig:vsLn2141}
\end{figure}

\begin{figure}
\includegraphics[scale=0.4]{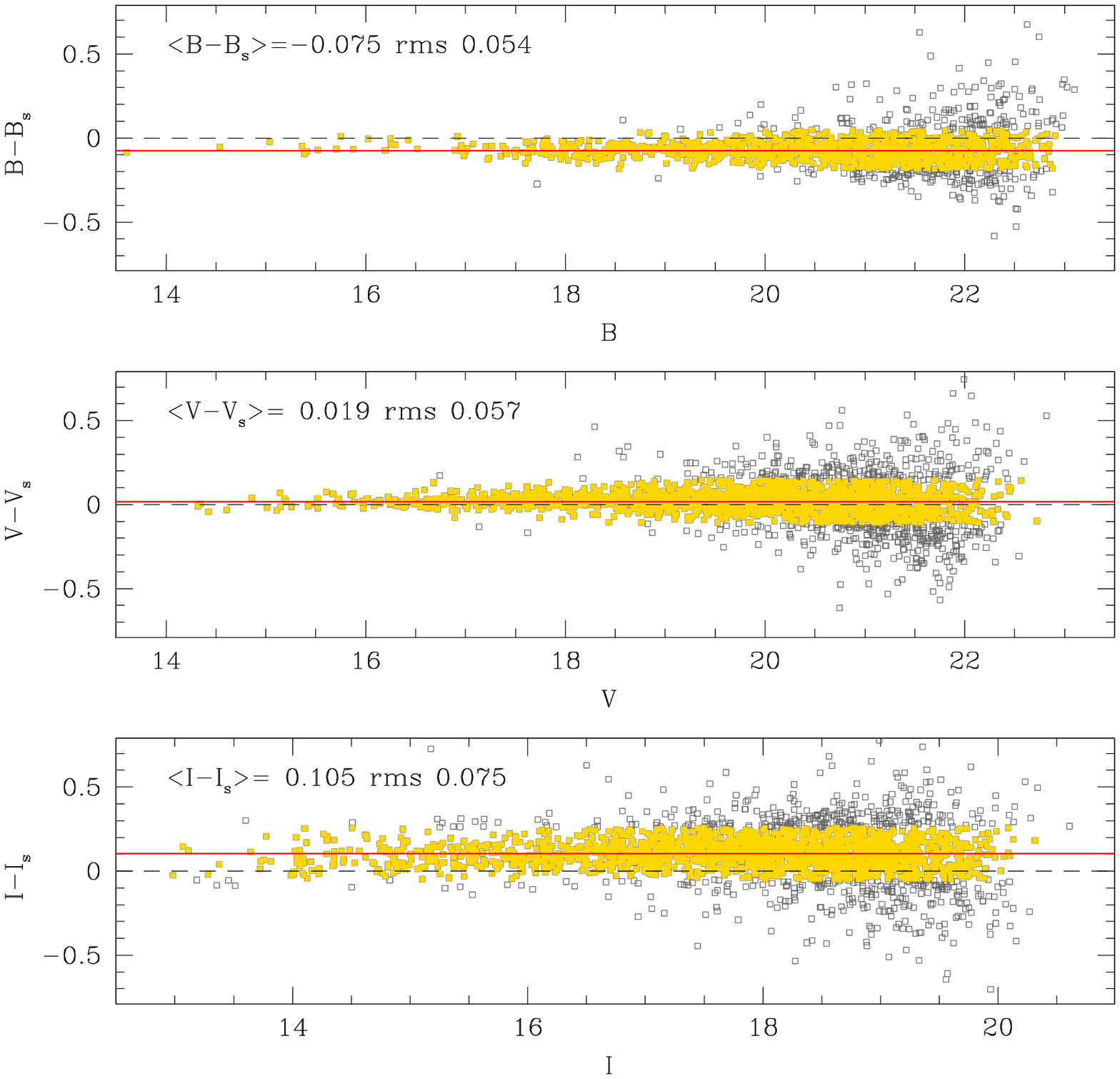}
\caption{Same as Fig. \ref{fig:vsLn1817}, but for Be81 compared to Sagar et al. (1998).}
\label{fig:vsLbe81}
\end{figure}

\begin{figure*}
\includegraphics[width=0.45\textwidth]{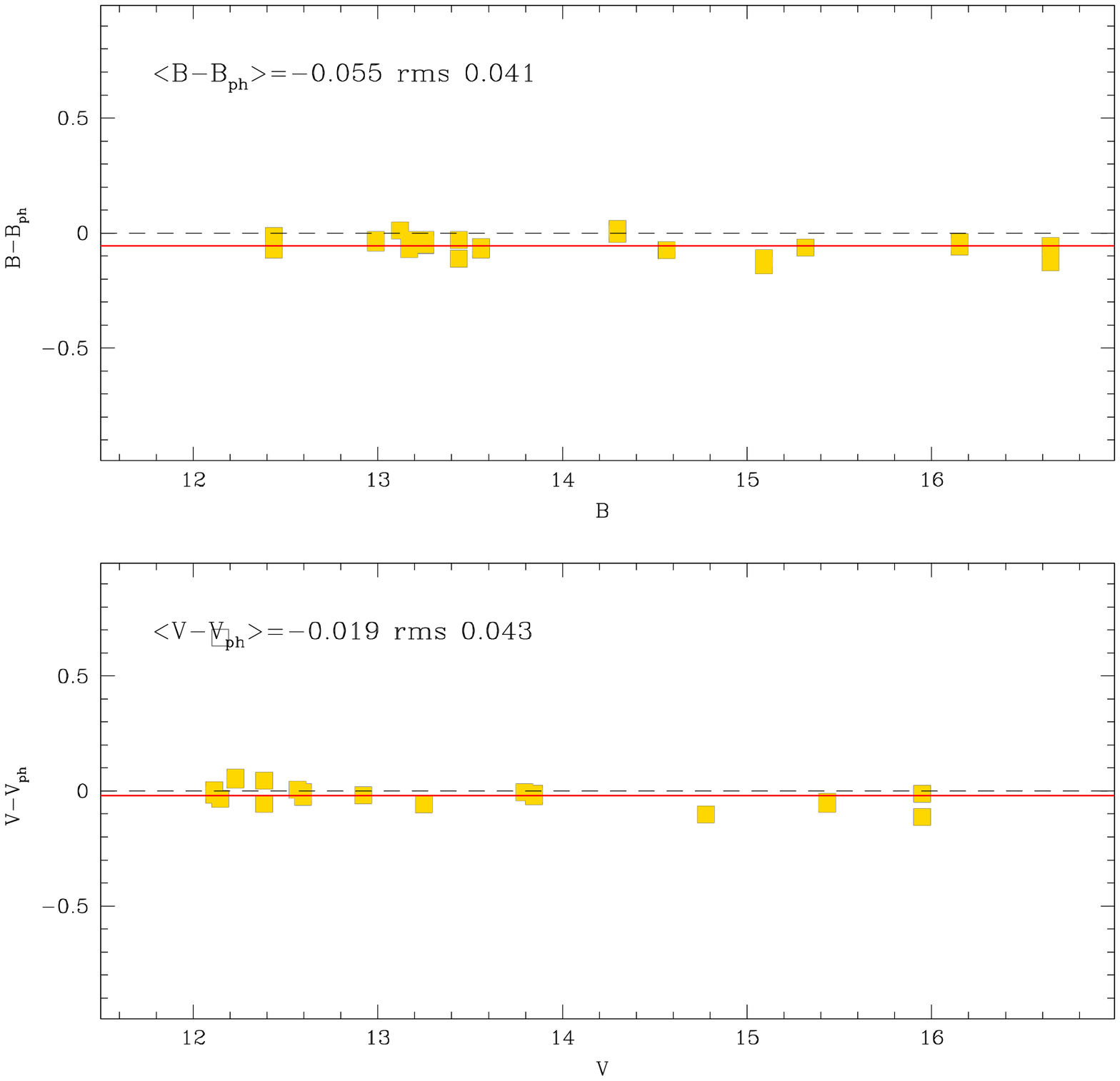}
\includegraphics[width=0.45\textwidth]{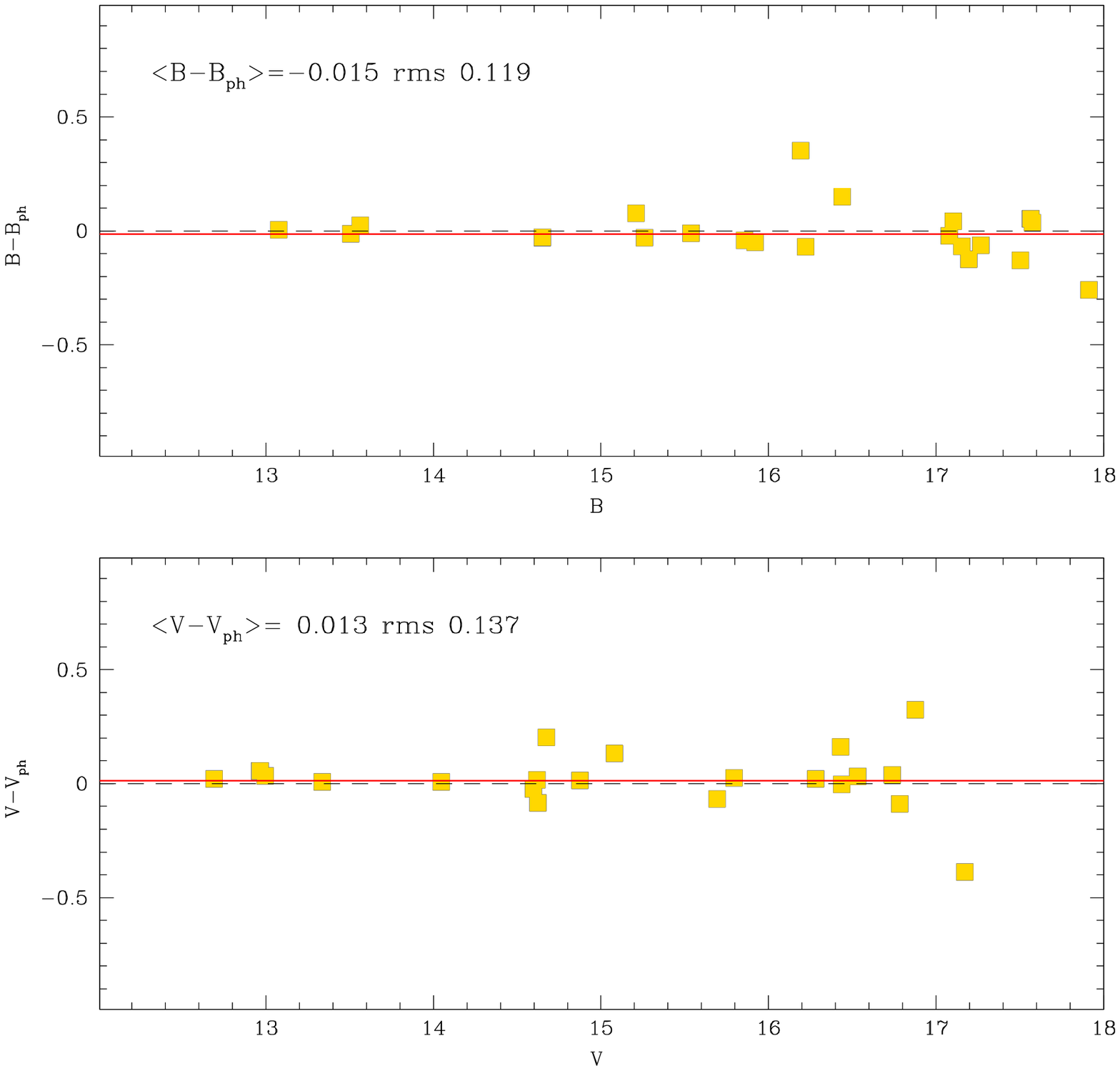}
\caption{Comparison of the $B$ and $V$ calibrated photometry with the photoelectric photometry. On the left the results for NGC~1817, on the right the case of NGC~2141.}
\label{fig:vsfotel}
\end{figure*}

In Figs. \ref{fig:vsLn1817}, \ref{fig:vsLn2141}, and  \ref{fig:vsLbe81} we show the comparisons of our calibration with the literature ones (downloaded through WEBDA\footnote{The WEBDA database is operated at the Department of Theoretical Physics and Astrophysics of the Masaryk University, see http://webda.physics.muni.cz}) for NGC~1817, NGC~2141, and Be~81.
In the case of NGC~1817, we find a small offset: about 0.04 mag in $B$ and 0.03 in $V$, corresponding to an offset of 0.01 mag in $B-V$. More worrisome are the comparisons obtained for NGC~2141 and Be~81, showing an offset of up to 0.1 mag. The explanation for such differences is not straightforward, since we can only perform relative comparisons, with no absolute reference point.  There must be issues
related to the adopted calibration equations, but it is not possible to identify in which data set. We have further investigated  this problem using photoelectric measurements, when available.  This was feasible for NGC~1817 and NGC~2141, thanks to the photoelectric data by \cite{harris77}, \cite{purgathofer64}, and \cite{burkhead72}, but not for Be~81. The results
 are shown in Fig. \ref{fig:vsfotel}. The agreement between our photometry and photoelectric standards is good for both clusters, showing only a tiny offset, smaller than 0.02 mag in most cases and only slightly worse for the $B$ of NGC~1817.

\begin{figure}
\includegraphics[scale=0.45]{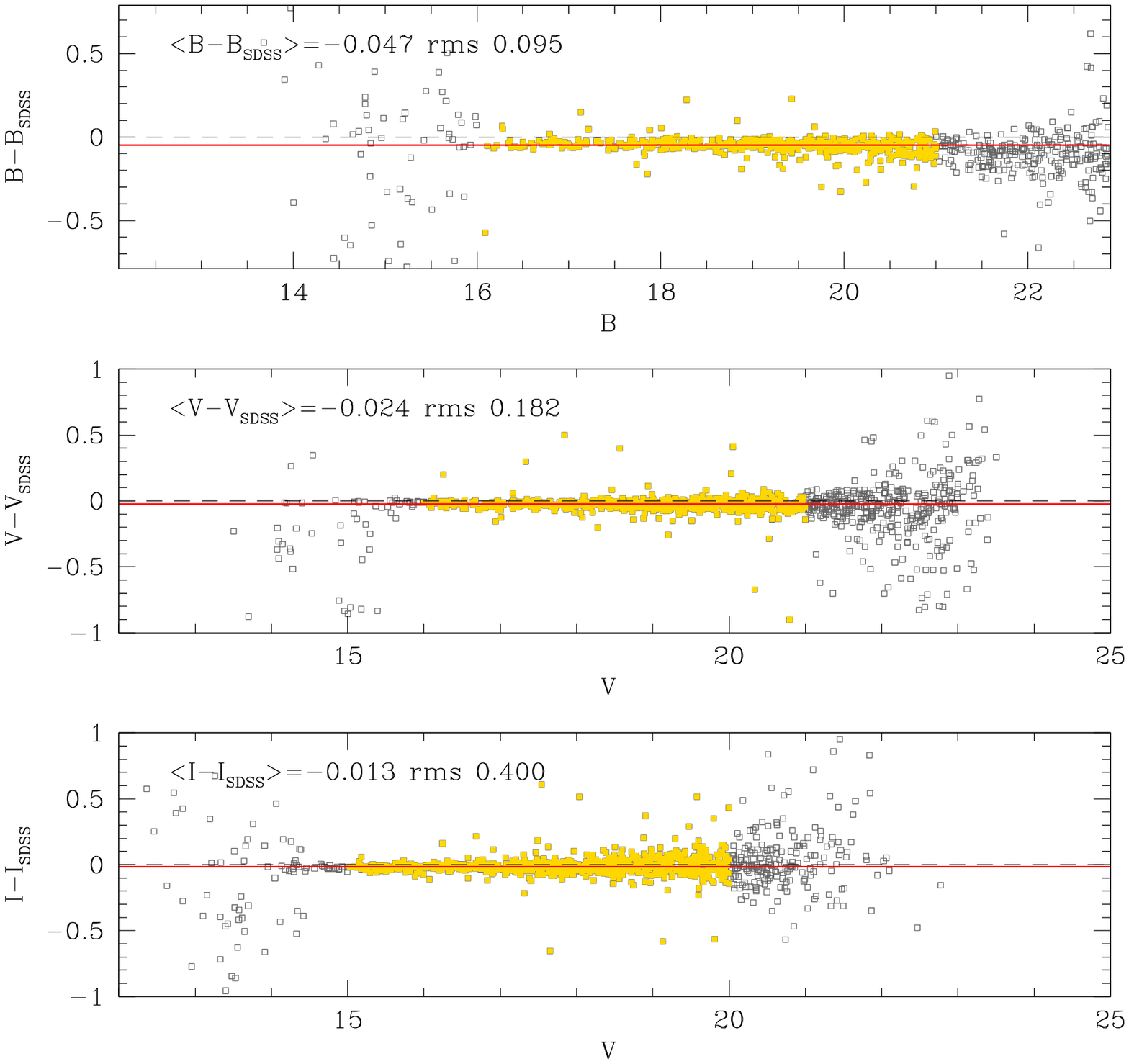}
\caption{Comparison of the $BVI$ with SDSS $ugri$ magnitudes calibrated to the Johnson-Cousins system for NGC~1817. The median of the difference (red line) is computed using the golden points: we excluded bright and possibly saturated stars and faint star, we used stars in common only with chip\#2 and flagged with $Q$ SDSS parameter equal to three.}
\label{fig:sdss1817}
\end{figure}

For NGC~1817 we were also able to compare our photometry with the Sloan Digital Sky Survey (SDSS)  \citep[see][]{york00} using the transformation by Lupton\footnote{http://www.sdss.org/dr4/algorithms/sdssUBVRITransform.html\#Lupton2005} to convert their magnitudes into the Johnson-Cousins system.
 The results are shown in Fig. \ref{fig:sdss1817}. The median of the difference in $B$ is about 0.05 mag, and is lower than 0.03 in $V$ and $I$. This translates in colour differences smaller than 0.03 mag.

In summary, we find that our photometry for NGC~1817 is in good agreement with the literature, and in particular  with both photoelectric measurements and SDSS data. 
For NGC~2141 we find a poor comparison with literature CCD data but a very good agreement with photoelectric measurements, which
makes us confident of our results. For  Be~81 there were no further checks feasible, but, given the robustness of the calibrations adopted for  the other two clusters,  we believe the third is correct too.

\subsection{The colour magnitude diagram}\label{sec:cmd}
\begin{figure*}
\includegraphics[scale=0.8]{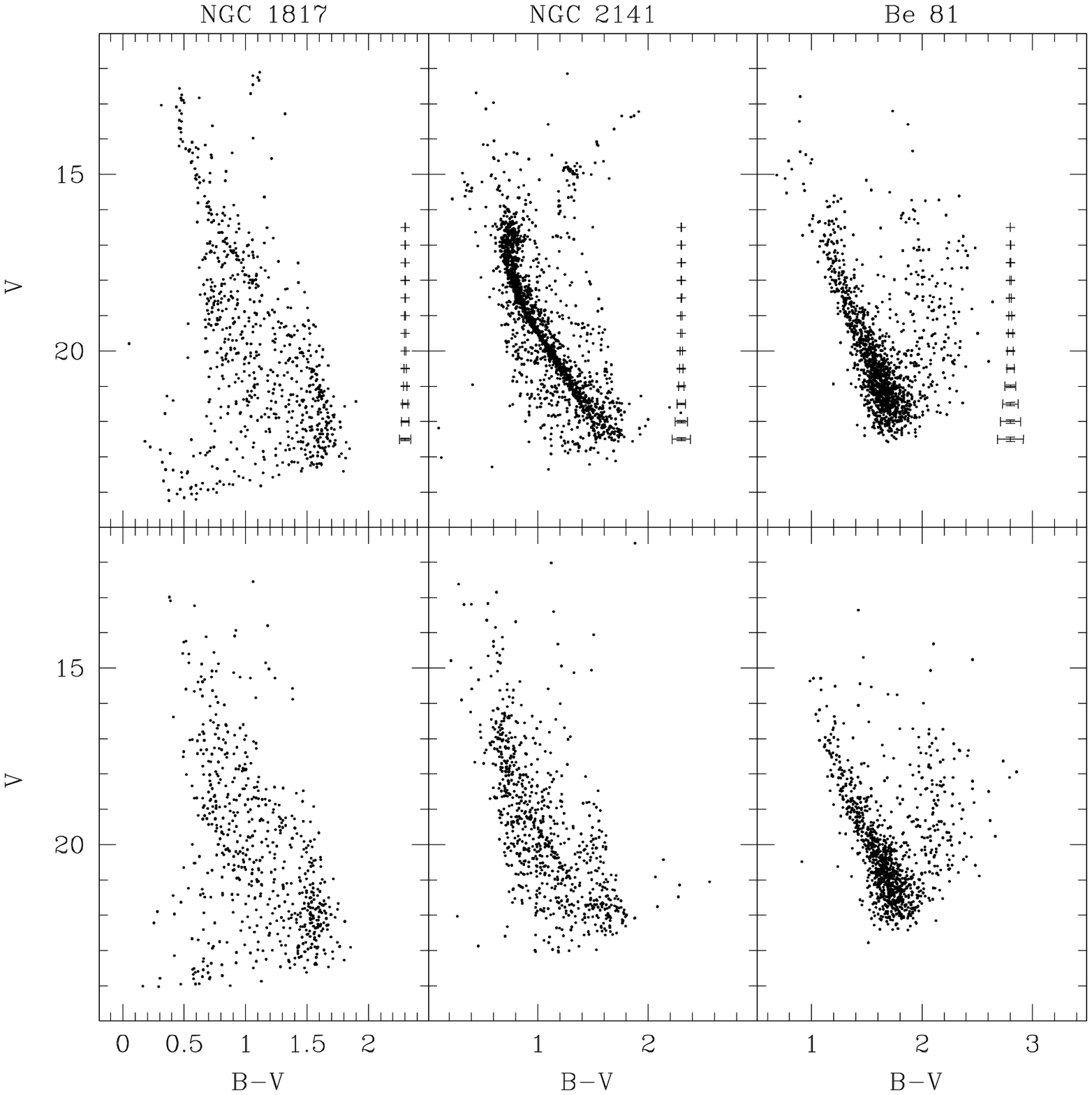}
\caption{\textit{Upper panels:} $V,B-V$ CMDs for the inner part of NGC~1817 (d$<5\arcmin$), NGC~2141 (d$<3\arcmin$), and Be~81 (d$<2.5\arcmin$). The errors on colour and magnitudes are indicated by error-bars and derived using the artificial stars tests. \textit{Lower panels:} $V,B-V$ CMDs for an external area with the same dimension.}
\label{fig:cmderrbv}
\end{figure*}

\begin{figure*}
\includegraphics[scale=0.8]{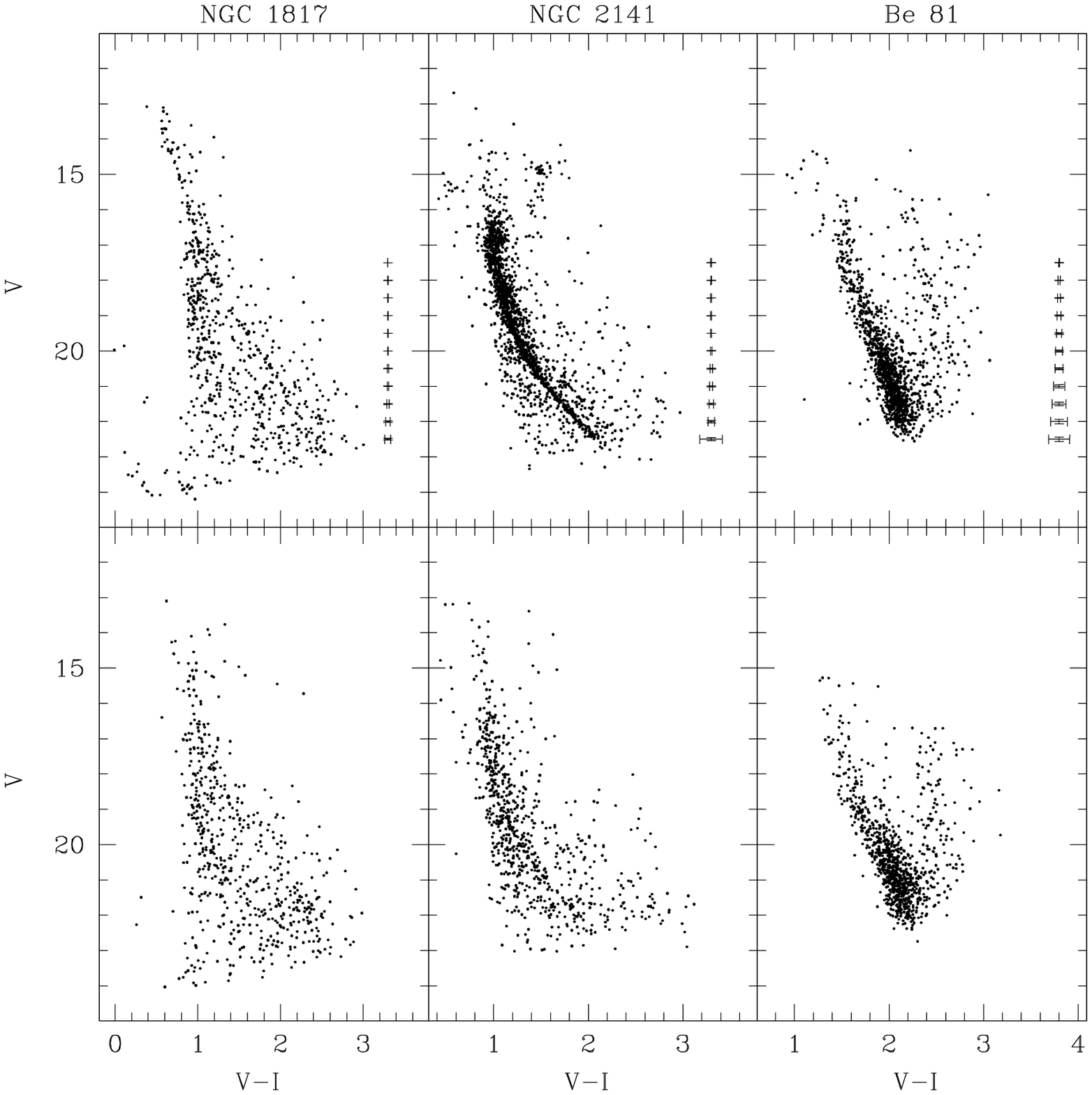}
\caption{As Fig. \ref{fig:cmderrbv}, but for $V,V-I$.}
\label{fig:cmderrvi}
\end{figure*}

The CMDs obtained for the three clusters are shown in Figs. \ref{fig:cmderrbv} and \ref{fig:cmderrvi}, with errors in colour and magnitude indicated. The errors are evaluated using the artificial stars tests. They are random standard errors, with no consideration of possible sources of systematics. In the upper panels only the more central regions are plotted, while more external regions are used for comparison to estimate the field contamination. For  Be~81 the
size  of the LBC FoV makes this possible, but NGC~2141 is present also in the outer parts of the FoV, and NGC~1817 is so extended that it fills all the four CCDs.

 The differences between the CMDs of the three OCs are quite evident. 

NGC~1817 is a young and luminous cluster,
but  not as rich as NGC~2141. Its size is probably larger than the LBT FoV, given the presence of probable cluster RC (at $V\sim12$ mag) and MS stars in the outer parts of our frames. The brighter MS and RC stars were saturated in I even in short exposures and we miss them  in the $V,V-I$ CMD.
 On the other hand we obtained a very good description of the MS, which extends for about 10 mag in $V$. 

NGC~2141 shows a very rich MS and a populated RC at $V\sim15$ mag. The RGB is visible at $B-V\sim1.6-2.0$  
up to $V=13$
 and there are a few probable sub giant branch (SGB) stars at its base. The binary sequence (redder and brighter than the MS) is very clear and neat. The TO is extended in colour, with a ``golf club'' shape common to other young OCs (see Sec. \ref{sec:diffredd}). A small clump of stars bluer and brighter than MS stars ($V\sim15.5$, $B-V\sim0.3$) is visible, probably blue-stragglers.

Be~81 is heavily contaminated by field stars,
and is hardly distinguishable, even using the control field for comparison.
However, there is a mild excess of stars at $V\sim16.5$ and $B-V\sim1.9$, which is not present in the outer field, and can be considered  the cluster signature, probably its RC.

The catalogue with the photometry of the three clusters will be made available through the CDS.

\subsection{Radial Velocity}\label{sec:RV}

\begin{table*}
\centering
\caption{Stars in common with Jacobson et al. (2009,2011) and Yong et al. (2005) in NGC~1817 and NGC~2141.}
\begin{tabular}{rcccccrrl}
\hline\hline
     ID & RA &Dec & $V$ &$B-V$ &$V-I$ &ID$_{webda}$ & RV &flag \\
\hline     
\multicolumn{9}{c}{NGC~1817 - stars in common with Jacobson et al. (2011)}\\
\hline
   1429 &  78.0807757 &  16.6801860 &  12.116 &   1.052 & 99.999 &     8 &   64.8 &     M \\
   1431 &  78.0260996 &  16.6376011 &  12.232 &   1.025 & 99.999 &    81 &   65.1 &     M \\
   1432 &  78.0190940 &  16.6740782 &  12.237 &   1.227 & 99.999 &    90 &   27.8 &    NM \\
   1434 &  78.0271406 &  16.7457104 &  12.389 &   1.048 & 99.999 &   177 &   65.2 &     M \\
   1435 &  78.0444686 &  16.6419862 &  12.494 &   1.027 & 99.999 &    79 &   65.8 &     M \\
   1436 &  78.0960056 &  16.5669407 &  12.480 &   1.268 & 99.999 &   155 &   14.5 &    NM \\
   1438 &  78.0772008 &  16.6959836 &  12.597 &   0.960 & 99.999 &    12 &   62.7 &     M \\
   1440 &  78.0941116 &  16.6357254 &  12.713 &   1.034 & 99.999 &    40 &   65.1 &     M \\
   1448 &  78.0939209 &  16.7331841 &  13.282 &   1.322 & 99.999 &    53 &   50.4 &    M? \\
   2922 &  78.1881838 &  16.5799475 &  13.808 &   0.844 &  0.976 &   219 &  -26.1 &    NM \\
   3106 &  78.1092441 &  16.5988226 &  12.079 &   1.039 & 99.999 &    72 &   66.5 &     M \\
   3108 &  78.1890040 &  16.7280692 &  12.106 &   1.115 & 99.999 &  2049 &   65.0 &     M \\
   3109 &  78.1498523 &  16.7244935 &  12.206 &   1.059 & 99.999 &    19 &   35.2 & M?,SB \\
   3110 &  78.2087521 &  16.6804613 &  12.253 &   1.100 & 99.999 &   127 &   65.1 &     M \\
   3111 &  78.1601438 &  16.7064247 &  12.336 &   1.110 & 99.999 &    22 &   63.7 &     M \\
   3112 &  78.2283968 &  16.6160611 &  12.361 &   1.126 & 99.999 &  2050 &   65.7 &     M \\
   3113 &  78.1356504 &  16.6660230 &  12.460 &   1.059 & 99.999 &    30 &   65.0 &     M \\
   3116 &  78.1717679 &  16.5846748 &  12.590 &   1.019 & 99.999 &   286 &   66.9 &     M \\
   3118 &  78.2083091 &  16.7333221 &  12.710 &   1.042 & 99.999 &   121 &   64.6 &     M \\
   3121 &  78.1291901 &  16.8236832 &  12.815 &   1.055 & 99.999 &   185 &   65.3 &     M \\
   3124 &  78.1221863 &  16.5986027 &  12.882 &   1.034 & 99.999 &    71 &   65.9 &     M \\
   3126 &  78.1834030 &  16.6199452 &  12.931 &   1.179 & 99.999 &   138 &    8.4 &    NM \\
   4510 &  78.3186789 &  16.6698094 &  13.663 &   1.066 &  1.192 &   471 &   48.1 &    NM \\
   4511 &  78.2917601 &  16.7518247 &  13.468 &   0.867 &  0.964 &  1722 &   15.1 &    NM \\
   4513 &  78.3164620 &  16.7476535 &  13.566 &   0.793 &  0.929 &   482 &   15.6 &    NM \\
   4518 &  78.3029299 &  16.7208041 &  13.736 &   0.881 &  0.955 &   477 &   40.3 &    NM \\
   4670 &  78.2593664 &  16.6553140 &  12.288 &   1.090 & 99.999 &   211 &   65.1 &     M \\
   6178 &  78.0818374 &  16.9064095 &  12.235 &   1.059 & 99.999 &  1292 &   65.5 &     M \\
\hline     
\multicolumn{9}{c}{NGC~2141 - $^a$Jacobson et al. (2009), $^b$Yong et al. (2005)}\\
\hline
6770 & 90.7115940 & 10.5078007 & 13.341 & 1.761 & -     & 1007$^a$ & 25.5 & M \\
     &  &  & & & & 1007$^b$ & 24.4 & M \\
6590 & 90.7427760 & 10.4441398 & 14.178 & 1.546 & 1.715 & 2066$^b$ & 24.8 & M \\
6604 & 90.7345371 & 10.4851441 & 14.777 & 1.385 & 1.606 & 1286$^b$ & 23.0 & M \\
6644 & 90.7511372 & 10.4788874 & 15.082 & 1.359 & 1.572 & 1333$^b$ & 23.5 & M \\
6771 & 90.7564588 & 10.4763049 & 13.337 & 1.871 & -     & 1348$^b$ & 24.6 & M \\ 
6776 & 90.7500858 & 10.5398554 & 14.081 & 1.500 & -     &  514$^b$ & 23.3 & M \\
6777 & 90.7814610 & 10.4469925 & 14.145 & 1.537 & -     & 1821$^b$ & 24.8 & M \\  
\hline
\end{tabular}
\label{tab:RV}
\end{table*}

\begin{figure}
\includegraphics[scale=0.45]{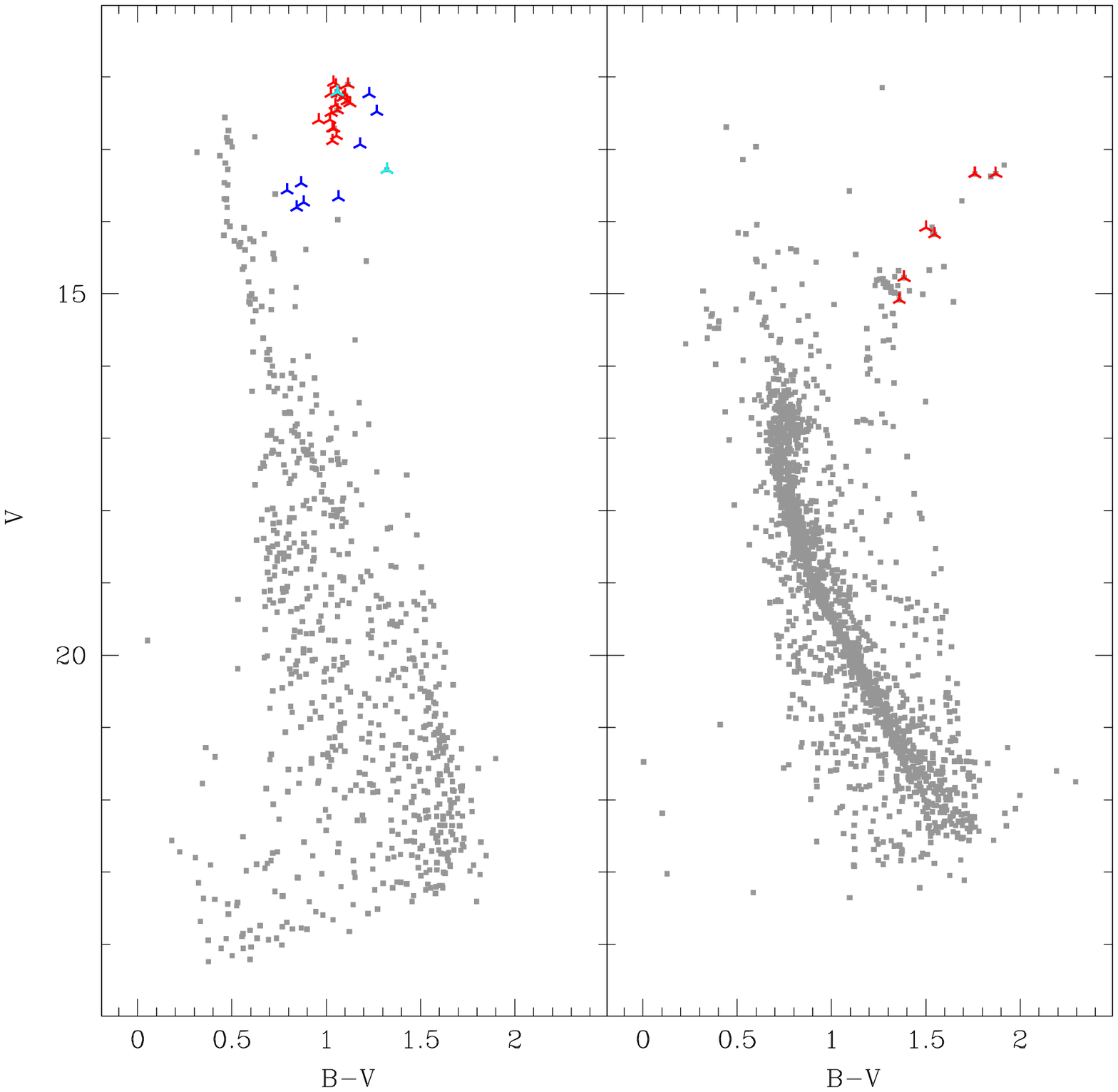}
\caption{\textit{Left panel:} CMD of NGC~1817 inside 5$\arcmin$. \textit{Right panel:} CMD of NGC~2141 inside 4$\arcmin$. The shaped points are the targets with RV measurements listed in Tab. \ref{tab:RV}. In red the sure members, in blue the non members, in cyan the stars with uncertain membership.}
\label{fig:RV}
\end{figure}

For NGC~1817 and NGC~2141,  we have identified the stars in our catalogue with literature RVs from  high-resolution spectroscopy. They are all evolved stars, mainly on the RC but also on the bright RGB. They are listed in Tab.~\ref{tab:RV}, and are displayed with larger symbols in the CMDs of Fig. \ref{fig:RV}.

For Be~81, \cite{wc09} observed stars in the CaT spectral region, but we
opted not to use 
their data because of the large uncertainty
in the membership attribution (see Introduction and their Sect. 3.1).

\section{Centre of gravity and density profile}
\label{sec:centre}
Exploiting the deep and precise photometry obtained with LBT and 
its large field of view, we re-determined the centre of each cluster following
 the approach described in \cite{donati12}. Briefly, we selected the 
 densest region on the images 
 by looking for the smallest coordinates interval that contains 70\% of all the stars. The centre is obtained as the average right ascension and declination when the selection is iterated twice. 
 For a more robust estimate, several magnitude cuts have been considered and the corresponding results averaged.  The root mean square (r.m.s.)
  on the centre coordinates is about 5$\arcsec$. 
  
  The most uncertain determination is for NGC~1817. It is a nearby cluster, hence its projected angular dimensions are 
  larger
 than the LBT's FoV. Moreover it is not richly populated and it does not seem particularly concentrated, circumstances that both hamper the analysis. We 
 thus applied the same method on the 2MASS catalogue to check the results on a larger field of view ($30\arcmin$ of radius).  We find a very similar answer, with a difference
  of only about half arcminute in both RA and Dec. We therefore adopted the value obtained from our photometry, which is  more precise and deeper than 2MASS, and allows us to include stars on the fainter MS.
The results are summarised in Tab. \ref{tab:OCcentre}.

\begin{table*}
\centering
\caption{Clusters centres and structural parameters. The rms on the centre determination is about 5$\arcsec$. }
\begin{tabular}{rcccccccc}
\hline\hline	
Cluster & RA$^a$ & Dec$^a$ & RA & Dec & $c$ & $r_c$ & $r_h$ & $r_t$	\\
  & (h:m:s) & ($\degr$:$\arcmin$ :$\arcsec$) & (h:m:s) & ($\degr$:$\arcmin$:$\arcsec$) &  & (arcsec) & (arcsec) & (arcsec) \\
\hline	
NGC~1817 & 05:12:15 & 16:41:24 & 05:12:38.33 & 16:43:48.85 & -   & -   & -   & -    \\ 
NGC~2141 & 06:02:55 & 10:26:48 & 06:02:57.71 & 10:27:14.43 & 1.0 & 120 & 234 & 1219 \\ 
Be~81    & 19:01:40 & -0:27:22 & 19:01:42.82 & -0:27:07.67 & 0.6 &  95 & 128 & 388  \\ 
\hline
\end{tabular}

$^a$Previous centre estimates from the web update of the \cite{dias_02} catalogue, see http://www.astro.iag.usp.br/$\sim$wilton/.
\label{tab:OCcentre}
\end{table*}
 
Following the approach adopted by \cite{cig_11}, the projected number density profile is 
determined by dividing the entire data-set in N concentric annuli, each one partitioned in four 
subsectors 
(although only two or three subsectors are used, if the available data sample only a portion of the annulus).
The number of stars in each subsector is counted and the density is obtained by dividing this value by the 
sector area. The stellar density in each annulus is then obtained as the average of the subsector densities, 
and the uncertainty is estimated from the variance among the subsectors. Also in this case, only stars within 
a limited range of magnitudes are considered in order to avoid spurious effects due to photometric incompleteness.
 
\begin{figure}
\includegraphics[bb=10 160 530 710,clip,scale=0.45]{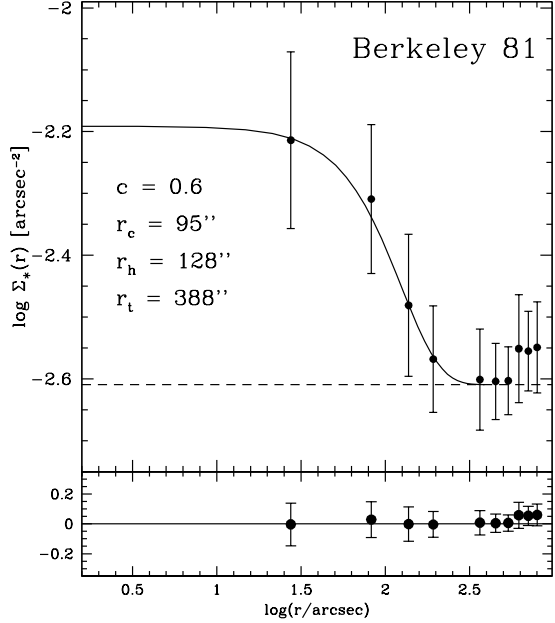}
\caption{King profile for Be~81.}
\label{fig:kbe81}
\end{figure}

\begin{figure}
\includegraphics[bb=10 160 530 710,clip,scale=0.45]{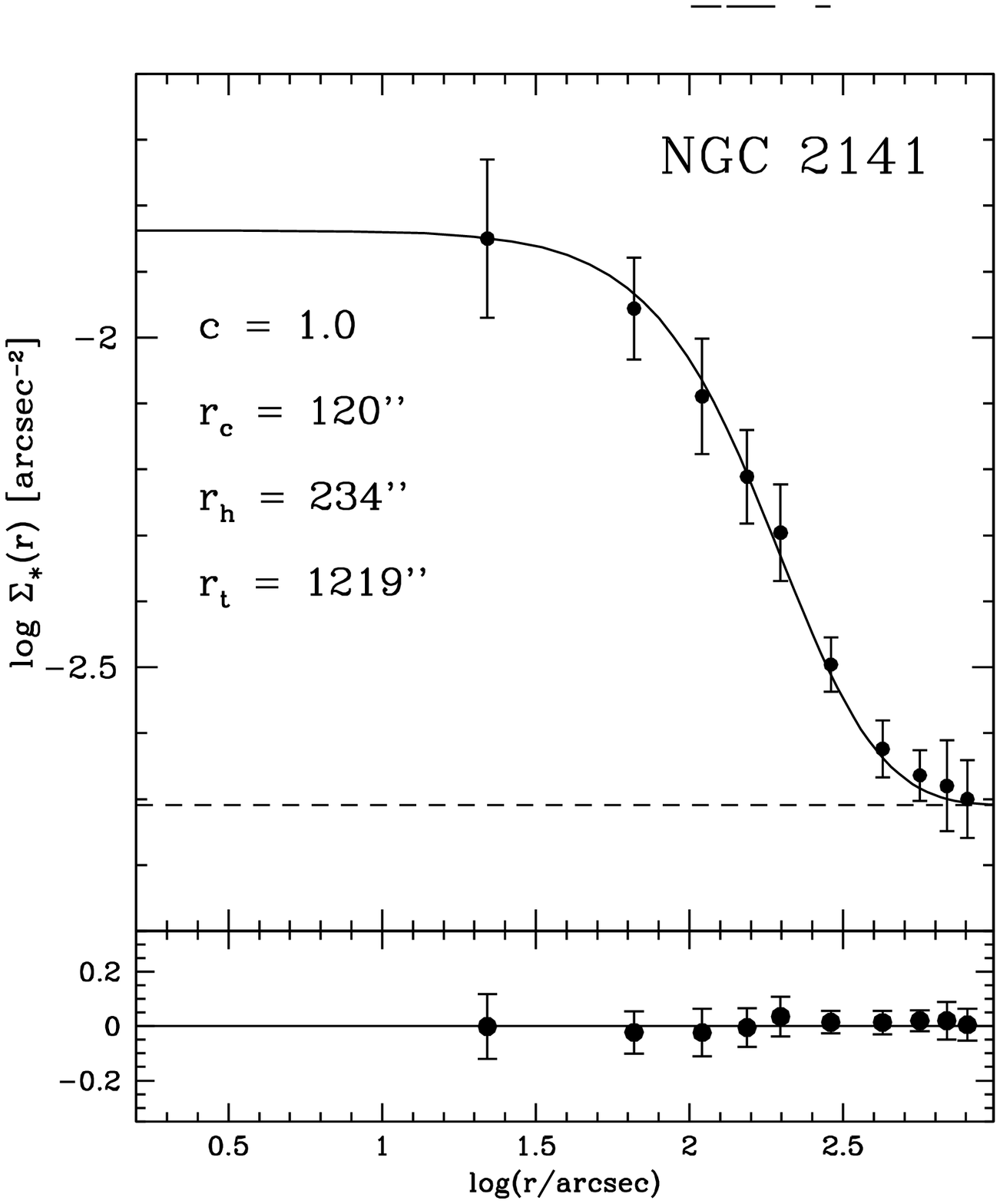}
\caption{King profile for NGC~2141.}
\label{fig:kn2141}
\end{figure}

The observed stellar density profiles are shown in Figs. \ref{fig:kbe81}, \ref{fig:kn2141} for the clusters
Be~81 and NGC~2141, respectively. For these two OCs the collected data-set covers the 
entire cluster extension, reaching the outermost region where the Galactic field stars represent the dominant
contribution with respect to the cluster. This is not the case for NGC~1817, which is not fully covered by the LBT's FoV. As done for the centre determination, we tried to evaluate its density profile on a larger area using 2MASS, SDSS, and literature catalogues, but the looseness of the cluster and its proximity to a nearby OC (NGC~1807, even if \citealt{balaguer04a} showed
that NGC~1807 is not a physical cluster) jeopardise the analysis. Unable to reach a satisfying conclusion, we preferred to limit the analysis to Be~81 and NGC~2141. The results are summarised in Tab. \ref{tab:OCcentre}.

In order to reproduce the observed profile, isotropic, single-mass King models \citep{king66} have been computed
adopting the~\citet[][]{sig95} code. The best
fit models are shown as solid curves 
 are shown in Figs. \ref{fig:kbe81}, \ref{fig:kn2141} together with  the observed density profiles.
In each figure we also show the values of concentration ($c=log_{10}(r_t/r_c)$), core radius ($r_c$), 
half-mass radius ($r_h$),
and tidal radius ($r_t$) as obtained from the best-fit model. The residual of the fit of the model to each observed point is shown in the lower panel of each density plot. 

Clearly, Be~81 is a small, low-mass
and very sparse OC. The density profile is hence affected by larger statistical uncertainty. Nevertheless, the residuals of the model fit  are quite small,
  at least in the most central part, where the
star counts are dominated by the cluster's 
 members.

\section[]{Differential reddening}\label{sec:diffredd}
As noted in Sec. \ref{sec:cmd}, NGC~2141 shows a ``golf club'' shaped MSTO. We can exclude that this observed feature is due to the photometric error, which is too small to explain the colour extension. \cite{carraro01} propose a metallicity spread as best explanation, but this circumstance is very unlikely in OCs. In literature there are other similar examples of  Milky Way OCs and Magellanic 
Clouds clusters showing an extended MSTO (see e.g. Tr~20 in the MW, \citealt{pla08}, and about 10 young globular clusters in the LMC, \citealt{mil09}). Another possible explanation is stellar rotation. For instance, \cite{bast09} find that fast rotators at the TO phase have a redder and fainter colour, and can be responsible
for  the ``golf club'' shape. \cite{gir11}, instead,
exclude that rotation can have such an effect. Also binary systems, which have redder colour and brighter magnitude than single stars, could explain the broadening of the MS, as could an age spread. 
The latter, however
 has never been convincingly observed in OCs. A more plausible explanation can be differential reddening (DR). Different absorptions on the cluster field due to different extinction paths along the line of sight results in different shifts in colour and magnitude. This circumstance can also explain the elongated shape of the RC, when RC stars are spread along one single direction. 

Most likely, DR is not negligible also over the field of Be~81, that 
is located very close to the Galactic plane (about 130 pc below the disc, see Sec. \ref{sec:CMDsynth}) and toward the Galactic centre. Its high average reddening, $E(B-V)\sim1.0$ mag, favours the chances for DR. However, Be~81 is severely contaminated by field stars, and this makes it  very hard to measure DR.
For NGC~1817 there is no direct evidence of DR from the observational CMD (see Fig. \ref{fig:cmderrbv}). 

To evaluate the effect of DR for NGC~2141 we adopt the following approach, using a revision of the method described in \cite{mil12}, adapted to the case of the OCs, which are less populated and more contaminated by field stars than the globular clusters. The main steps of the process are the following: 
\begin{itemize}
\item we draw a fiducial line along the MS, and use it as a reference locus for the DR estimate;
\item we draw a box on the MS: all stars falling in this box are used to estimate the DR. The box is chosen to select stars on the blue side of the MS, and to avoid as many binaries as possible, since they also produce a shift to the red of the sequence. We also keep far from the MSTO and the fainter part of the MS, where errors are larger and field stars confuse the picture;
\item for each star in the catalogue we pick the 30 nearest and brightest stars inside the MS box and compute their median distance along the reddening vector direction from the fiducial line in the CMD plane. This distance is used to correct colour and magnitude for DR;
\item after the correction for the first DR estimate is applied star-by-star, the algorithm starts a new loop and this procedure is repeated until a convergence is reached. The convergence criterion is a user-defined percentage of stars for which the DR correction is lower than the average rms on these estimates; 
\item once a final value for the DR is obtained for each star, a binning is performed in the spatial plane. The spatial scale 
must be 
compatible with the average distance of the 30 neighbour stars selected and used for the DR estimate. In our case it is less than 1 arcmin$^2$, as described in the following paragraphs. At this point 
the outliers are rejected, i.e., stars whose DR estimate is larger
 than the average error, and stars whose  distance to the 30 neighbours is larger than average.
\item a final and robust value for the DR is then computed as the average value of the DR corrections associated to the stars falling in the same bin and the error on this estimate is the associated rms. The values obtained are not absolute values but relative
 to the fiducial line.  
\end{itemize}

\begin{figure}
\begin{center} 
\includegraphics[scale=0.45]{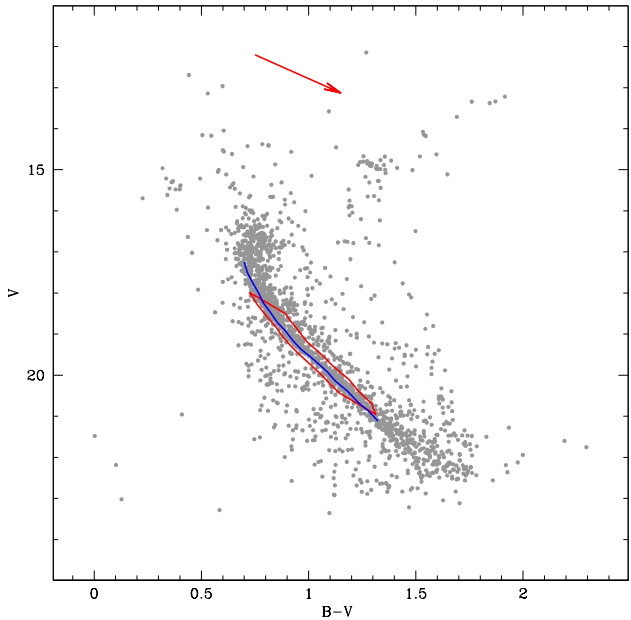} 
\caption{\label{fig:cmd_box}CMD of NGC~2141 inside 4 arcmin. The red box and the blue line indicate the MS box and the fiducial line for the DR estimate. The red arrow indicates the reddening vector.} 
\end{center}
\end{figure}

We estimated the DR in the $B-V$ colour. The direction of the reddening vector is derived assuming the standard extinction law ($R_V=3.1$, $E(V-I)=1.25\times E(B-V)$) described in \cite{dean_78}. The fiducial line is defined using the CMD of the inner part of the cluster (all the stars inside 4$\arcmin$) and is chosen as the ridge line along the MS. Several attempts have been made to avoid fiducial lines that, during the estimates of the DR, lead to corrections that artificially and significantly change the magnitude and colour of the age-sensitive indicators (e.g. RC, MSTO). We 
want in fact to keep RC, MSTO and the blue envelope of the MS as close as possible to the original position in the CMD, to avoid
spurious interpretations of the cluster parameters due to DR corrections. When defining the MS box we avoided the broad and bended region of the TO, where the morphology could hamper the correct interpretation, and the fainter part of the MS, where the photometric error is more important. The box and the fiducial line used are highlighted in Fig. \ref{fig:cmd_box} with colours.

Taking into account the star counts of the inner and outer parts of the cluster (see Sec. \ref{sec:centre}) we decided to limit the 
DR correction  to stars within a 4$\arcmin$
radius (approximately the half mass radius). For the outer 
regions
the contamination of field stars becomes not negligible (the contrast density counts with respect to the field plateau drops below 50\%) and any attempt to estimate the DR is severely affected by field interlopers. The spatial smoothing applied to have a more robust statistic is $0\arcmin.4\times 0\arcmin.4$ in right ascension and declination. As final caveat, we stress that photometric errors, undetected binary systems, and residual contamination from the field could affect the DR estimation, since they all produce a broadening of the MS. Our results are then an upper limit to the DR. 

\begin{figure}
\includegraphics[scale=0.45]{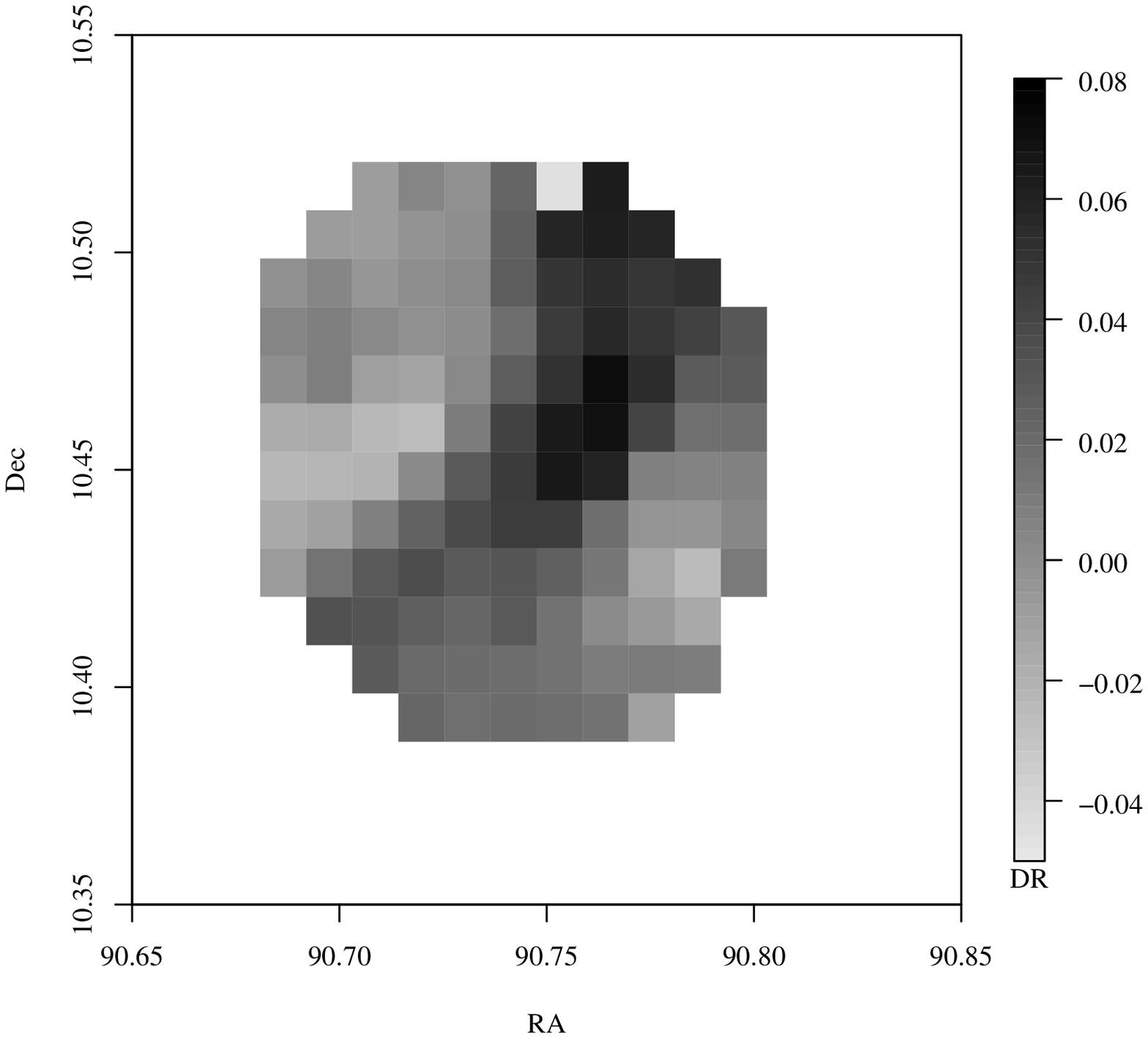}
\includegraphics[scale=0.45]{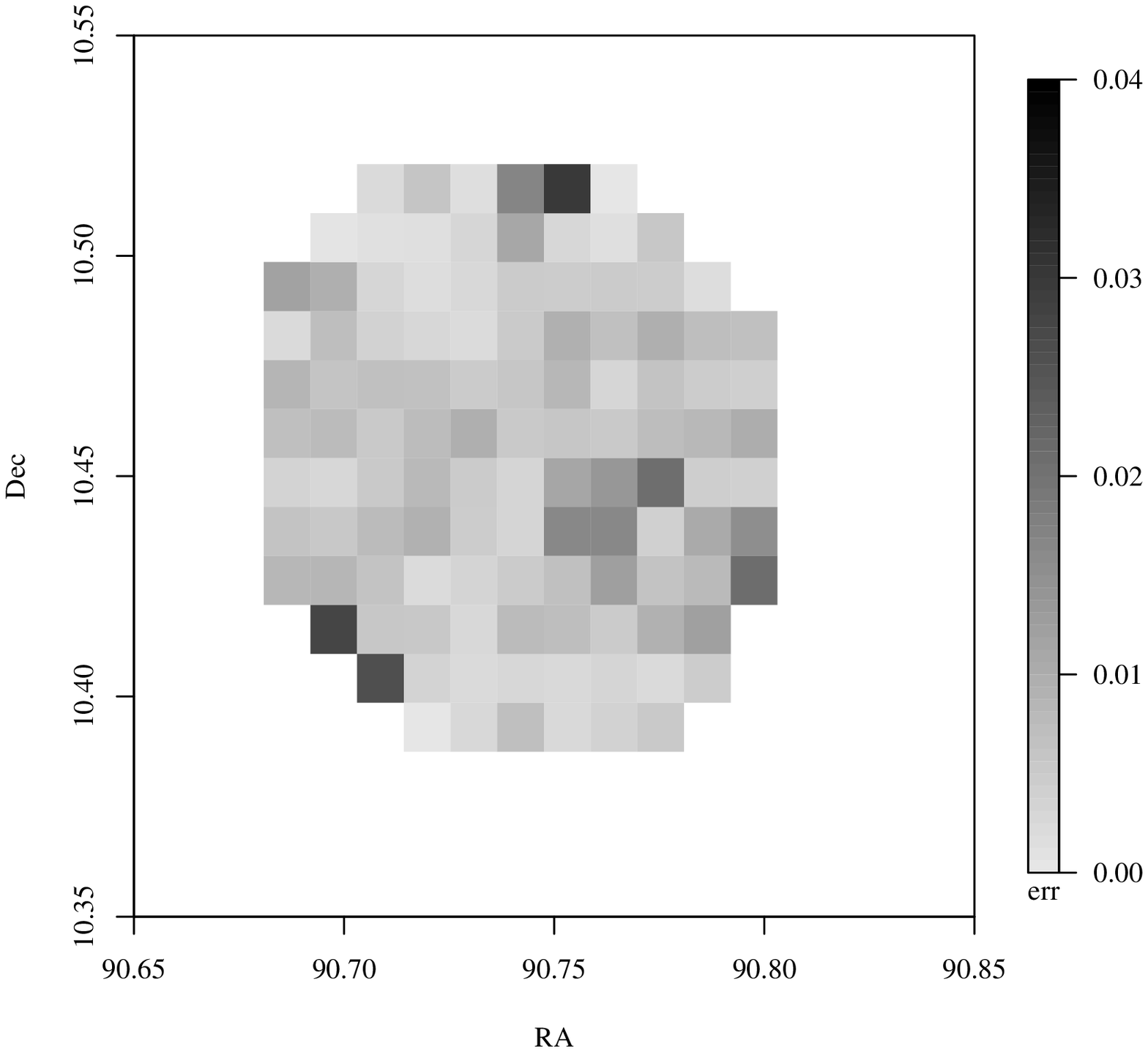}
\caption{\textit{Top panel:} Colour deviations from the reference line due to the effect of DR, mapped on a $0\arcmin.4\times0\arcmin.4$ grid for stars inside 4$\arcmin$ from the centre. The correction is expressed in gray-scale colours, see the legend on the right side. \textit{Bottom:} Corresponding error map.}
\label{fig:drgrid}
\end{figure}

In Fig. \ref{fig:drgrid} we show the map of the DR obtained in terms of $\Delta E(B-V)$ with respect to the fiducial line. It ranges from $\sim-0.04$ to $\sim+0.1$. In the same figure we show the corresponding map of the error associated to our estimates. The discrete appearance of these maps is due to two 
reasons:  the poor sampling of a circular area with
 polygonal bins and the avoidance of interpolation in the corners, where the poor statistics could lead to uncertain estimates.

\begin{figure}
\includegraphics[scale=0.45]{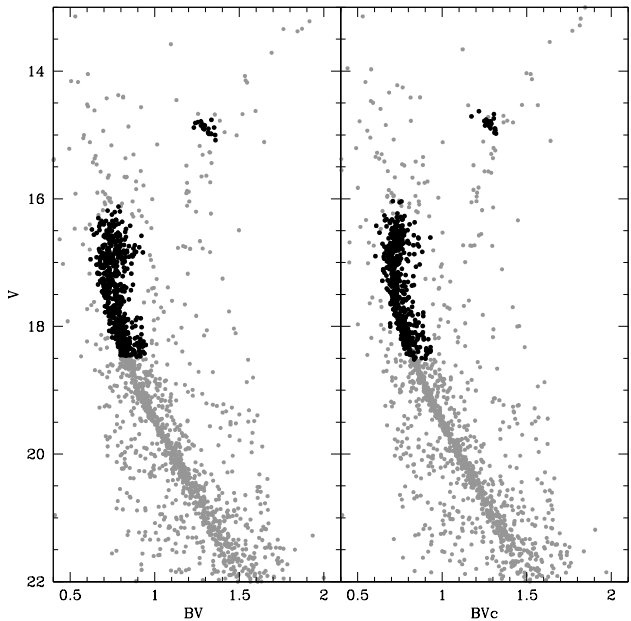}
\caption{CMDs for NGC~2141 inside 4 arcmin. \textit{Left panel:} observational CMD. \textit{Right panel:} CMD after correction for DR. The stars in the upper part of the MS and in the RC phase are highlighted in black to better show the effect of the correction.}
\label{fig:drn2141}
\end{figure}

The overall effect of the DR correction on the CMD appearance is shown in Fig. \ref{fig:drn2141}. The MS and MSTO region appear tighter, reducing substantially the broadening. 
In the figure, only the upper MS stars corrected for DR are highlighted in black,  but  the lower MS benefits from the DR correction too.
 The RC stars, 
 apparently  aligned along the direction of the reddening vector in the original
  CMD (see the left panel in Fig. \ref{fig:drn2141}), appear more clumped after the DR correction, thus supporting the DR hypothesis. 
  Also the RGB looks better defined. Furthermore, our DR estimate 
  does   not change the luminosity level and colour of age sensitive indicators such as the MSTO, or the bright edge  of the MS, the red-hook phase. 
  We list in our catalogue for NGC~2141 both the original magnitudes and the DR corrected ones.

We cannot apply the same analysis to
 Be~81 because it is severely contaminated by field interlopers even in the inner regions.
 For instance, in the central  2$\arcmin$ (approximately the half-mass radius estimated in Sec. \ref{sec:centre}), the density contrast is only 25\%.
 Hence, in its  case the algorithm  would be driven
  by stars not belonging to the cluster rather than MS stars, seriously weakening the results.
 We thus prefer to evaluate the effect of DR on Be~81 with the synthetic CMD technique described in the next section.

\section[]{Synthetic CMD}
\label{sec:CMDsynth}

Age, metallicity, distance, mean Galactic reddening, and binary fraction have been estimated
with the same procedure adopted for other works of this series \citep[see][and references therein]{donati12,ahumada13}. We compare the observational CMDs with a library of synthetic ones, built using synthetic stellar populations \citep[see e.g., ][]{cig_11}. Different sets of evolutionary tracks\footnote{The Padova \citep{bre_93}, FRANEC \citep{dom_99}, and FST ones \citep{ven_98} of all available metallicities, as in all the papers of the BOCCE series.}
have been used to Monte Carlo generate the synthetic CMDs. The best fit solution is chosen as the one that can best reproduce some age-sensitive indicators as the luminosity level of the MS reddest point (``red hook'', RH), the RC and the Main Sequence Termination Point (MSTP, evaluated as the maximum luminosity reached after the overall contraction, OvC, and before the runaway to the red), the luminosity at the base of the red giant branch (RGB), the RGB inclination and colour, and the RC colour.
The most valuable age indicators are the Turn Off (TO) point, that is the bluest point after the OvC, and the RC luminosity; however, at least in the case of OCs, these phases may be very poorly populated, and identifying them is not a trivial game, especially if a strong field stars contamination is present (as in the case of Be~81).

The binary fraction is estimated adopting the method described in \cite{cig_11}. The DR is taken into account and the synthetic CMD technique applied to the 
DR corrected photometry. The best fit to all the above indicators provides the best choice for age, reddening, and distance modulus. To infer the metallicity it is crucial to analyse together 
all the $BVI$ photometry  \citep[see][]{tosi_07}:  the best metallicity is the one that allows to reproduce at the same time both the observed $B-V$ and $V-I$ CMD.
To deal with $(B-V)$ and $(V-I)$ colours we adopted the normal extinction law \citep{dean_78}.

We estimated the errors on the cluster parameters considering both the instrumental photometric errors and the uncertainties of the fit analysis, as done in \cite{donati12}. The net effect of the former is an uncertainty on the luminosity level and colour of the 
adopted indicators. This affects mainly the estimate of the mean Galactic reddening and distance modulus, as they are directly defined by matching the level and colour of the upper MS and the RH and MSTP indicators.
 We must also consider the dispersion in the results arising from the fit analysis. Open Clusters offer poor statistics, and important indicators, such as the RC locus, are poorly defined. Hence, we cannot find a unique solution, but only a restricted range of viable solutions. In practice, we select the best fitting synthetic CMD and then take into account the dispersion of the cluster parameters estimates in the error budget. 
 The uncertainties are assumed
to be of the form:
$$\sigma^2_{E(B-V)}\sim\sigma^2_{(B-V)}+\sigma^2_{fit}$$
$$\sigma^2_{(m-M)_0}\sim\sigma^2_{V}+R_V^2\sigma^2_{E(B-V)}+\sigma^2_{fit}$$
$$\sigma^2_{age}\sim\sigma^2_{fit}$$
Typical photometric errors are $\sim0.04$ on the reddening and  $\sim0.1$ on the distance modulus (assuming negligible the error on $R_V$). The  error resulting from
 the fit analysis depends mainly on the uncertainty on the RC level and on the coarseness of the isochrone grid. It is of the order of $\sim0.02$ for the reddening, and ranges between 0.01 and 0.05 for the distance modulus, and about 0.2-1 Gyr for the age.

\subsection{NGC~1817}
\begin{figure}
\includegraphics[scale=0.45]{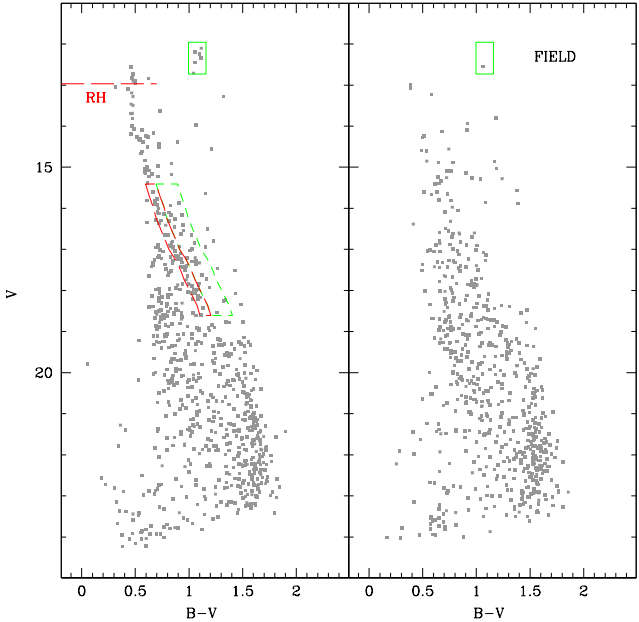}
\caption{{Left panel:} $V,B-V$ CMD for the inner part of NGC~1817 (inside 5$\arcmin$). The age indicators RC (green box) and RH (red line) are shown. The red and green boxes on the MS and redward of it are used to estimate the percentage of binaries. \textit{Right panel:} CMD of the comparison field of the same area. The same RC box adopted in the left panel is shown here.}
\label{fig:n1817cmd}
\end{figure}

With the deep LBT photometry  we can reach magnitude $V\sim23$ in the $B-V$ CMD, describing very well the MS. The RC is well visible at $V\simeq12.3$, as shown in Fig. \ref{fig:n1817cmd}. In the same figure we show the comparison with an external 
region of the same area. 
We can see the signature of the cluster 
(mainly MS stars) also in the outer parts  of the image.
 As explained in Sec. \ref{sec:centre}, we could not cover the whole extension of the cluster with the instrument's FoV. There is a clear signature of RC stars, confirmed by the studies on the RV of spectroscopic targets (see Sec. \ref{sec:RV} and Fig. \ref{fig:RV}); the upper part of the MS is poorly populated so it is difficult to reach a statistically firm conclusion on the locus of the RH. We place this phase at magnitude $V\simeq13$. A well defined binary sequence is visible redward of the MS.

To estimate the binary fraction we defined two CMD boxes, one which encloses MS stars and the other redward of the MS in order to cover the binary sequence (see dashed lines in Fig.~\ref{fig:n1817cmd}). To remove the field contamination we subtracted the contribution of field stars falling inside the same CMD boxes 
in a portion of the control field with same area.
 We performed the same computation on regions smaller and larger than 5$\arcmin$, finally ending with an estimate between 20\% and 30\%. The dispersion on the estimate is mostly due to the spatial fluctuations across the control field. For example in the inner 
  area around the cluster centre a higher fraction of binaries is found. 
  Notice that the derived binary fractions may be underestimated,  since  we possibly miss systems with very low mass secondary, whose luminosity doesn't alter significantly that of the primary. A mean fraction of 25\%  has been assumed for all the simulations presented here.

We limit the differential reddening to 0.02 mag because we find no direct evidence of it in this cluster.

After fixing these two parameters we use the synthetic CMD technique to estimate the age, reddening, and distance modulus of the cluster. For the simulations we used all the stars inside $5\arcmin$ from the centre.

Using the Padova models we find that a subsolar metallicity is required to describe with the same model both the $V,B-V$ and $V,V-I$ observational CMDs. In particular the best match is obtained for $Z=0.008$ ([Fe/H]$\simeq$-0.40), an age of 1.1 Gyr,  $E(B-V)=0.23$, and $(m-M)_0=11.1$.

In the case of the FST models we converge to similar results, finding the best solution for a metallicity lower than solar. We 
chose $Z=0.01$, an age of 1.05 Gyr, $E(B-V)=0.21$, and $(m-M)_0=10.98$.

For the FRANEC models we find the best fit for $Z=0.01$, age of 0.8 Gyr, $E(B-V)=0.34$, and $(m-M)_0=10.88$. The age is younger 
than with  the other two models, as expected since these evolution tracks
do not include overshooting. We reproduce the magnitude of the age sensitive indicators (RH and RC), but we don't match the RC colour and the MS shape and colour. In particular, the FRANEC models
 cannot reproduce the correct inclination of the MS for $V>16$. 

\begin{figure}
\includegraphics[scale=0.45]{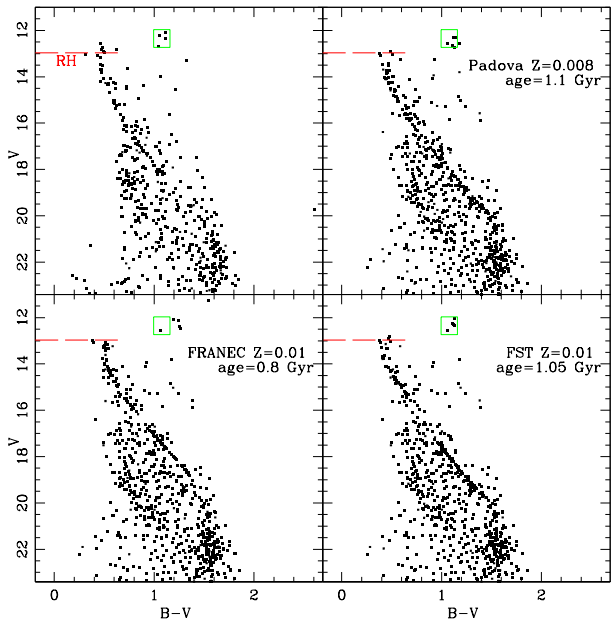}
\caption{Top left panel: CMD of stars inside 5$\arcmin$ radius area of NGC~1817. Clockwise from the top right panel: the best fitting synthetic CMD obtained with Padova, FST, and FRANEC models.}
\label{fig:n1817synth}
\end{figure}

Fig.~\ref{fig:n1817synth} shows the comparison between the observational CMD (top left) and the best fits obtained with the three sets of tracks.

\begin{figure}
\includegraphics[scale=0.45]{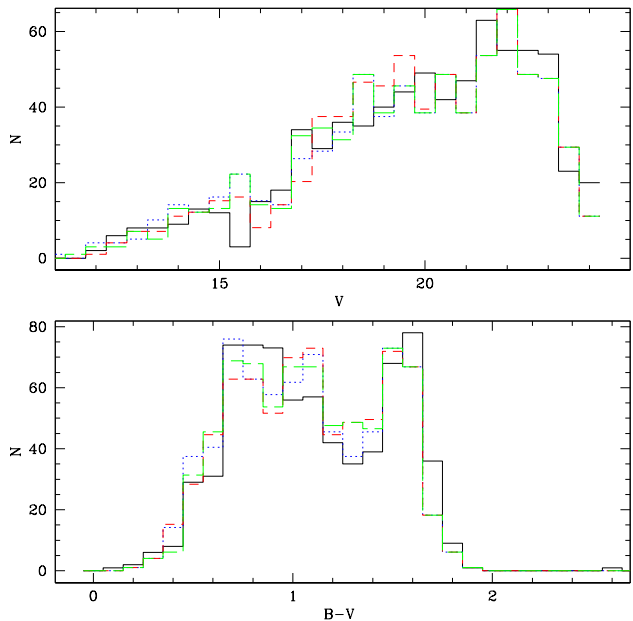}
\caption{Luminosity functions in magnitude $V$ (upper panel) and colour $B-V$ (lower panel). The solid black line is obtained from the observational CMD, the blue dotted line from the Padova synthetic CMD, the red dashed line from the FST synthetic CMD, and the green dot-dashed line from the FRANEC synthetic CMD.}
\label{fig:n1817lfs}
\end{figure}

The luminosity functions (LFs, see Fig. \ref{fig:n1817lfs}) show a satisfying agreement. There are small departures between the observational and synthetic LFs probably due to the poor statistics in star counts. For example the observational CMD (Fig. \ref{fig:n1817cmd}) shows a lack of stars at $V\sim19$ which is not reproduced in any synthetic CMDs.

From this analysis it turns out that the Padova and FST models provide a better description of the observational CMDs. This restricts the best age to 1.05-1.1 Gyr. Consequently the Galactic reddening is about 0.22\footnote{The \cite{sch_98} estimate is 0.43 mag, but this is the asymptotic value in that direction, while the cluster is nearby},
 while the distance modulus is between 10.98 and 11.1. The results are summarised in Tab. \ref{tab:summary}. 

\cite{balaguer04a} estimate an age of about 1.1 Gyr, a reddening of 0.21 and a distance modulus of 10.9. 
Our results are in excellent agreement with theirs.
The metallicity of the cluster is well defined by several high-resolution spectra analysis, and different works show very similar results of about [Fe/H]$\simeq$-0.34 (see Introduction). 
Our photometric analysis suggests a metallicity ranging from -0.40 to -0.30 and confirms these findings.  

\subsection{NGC~2141}
\begin{figure}
\includegraphics[scale=0.45]{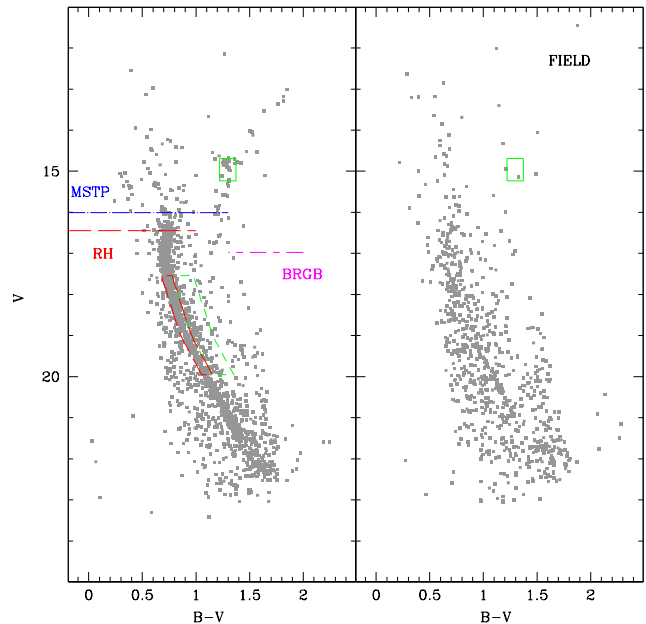}
\caption{ \textit{Left panel:} $V,B-V$ CMD for the inner part of NGC~2141 (inside 4$\arcmin$) corrected for DR. The age indicators RC (green box), RH (red line), MSTP (blue line), and BRGB (magenta line) are shown. The red and green boxes on the MS and redward the MS are used to estimate the percentage of binaries. \textit{Right panel:} CMD of the comparison field of the same area. The cluster is still visible, even if as a minor component.}
\label{fig:n2141cmd}
\end{figure}

NGC~2141 shows clearly all its evolutionary sequences. In Fig. \ref{fig:n2141cmd} we show the comparison of the inner part of the cluster (inside 4$\arcmin$, corresponding to the half-mass radius of the cluster) with an external 
region of the same area.  Even in the outer parts of the instrument FoV the cluster it is clearly present, 
with an evident star excess at $V\sim20$ aligned along the MS direction, and a mild excess at brighter magnitudes. 
We identify
 the RH at $V\simeq16.4$, the MSTP at $V\simeq16$, and the RC at $V\simeq15$ and $B-V\simeq1.3$. We find an indication of stars in the SGB phase at the base of the RGB and identify the BRGB at $V\simeq17$.

We evaluated the fraction of binaries as for NGC~1817, and find
 an average fraction of 16\%. For the simulations we use the photometry corrected for DR (see Sec. \ref{sec:diffredd}), 
 and adopt a DR of
  0.02 mag to take into account the intrinsic scatter in the correction. 

Keeping fixed these parameters we estimate the cluster age and metallicity comparing the observational CMD for stars inside $4\arcmin$ 
from the cluster centre
with our synthetic CMDs. We find that only models with metallicity $Z<0.02$ are in agreement with both $(B-V)$ and $(V-I)$, therefore we discard models with solar metallicity. 

For the Padova models we obtain the best match using the metallicity $Z=0.008$ ([Fe/H]$\sim$-0.4). Our synthetic CMD reproduces the magnitude and colour of all the age indicators,
reproducing very well the MS, the binary sequence, and the RGB. The corresponding cluster parameters are: age 1.9 Gyr, $E(B-V)=0.36$, and $(m-M)_0=13.2$.

With the FST models we find a good match for $Z=0.006$, age 1.7 Gyr, $E(B-V)=0.45$, and $(m-M)_0=13.06$. Also in this case the synthetic CMDs can reproduce well the MS, the binary sequence, and the RGB even if the RC colour is slightly redder than observed.

In the case of the FRANEC models, the best fit is obtained for $Z=0.01$, age of 1.25 Gyr, $E(B-V)=0.45$, and $(m-M)_0=13.19$. Despite being able of 
matching the luminosity of the age sensitive indicators, the colour of one of them, the RC, is much redder than observed. Moreover, the MS shape is poorly reproduced for faint magnitudes ($V>19$). 

\begin{figure}
\includegraphics[scale=0.45]{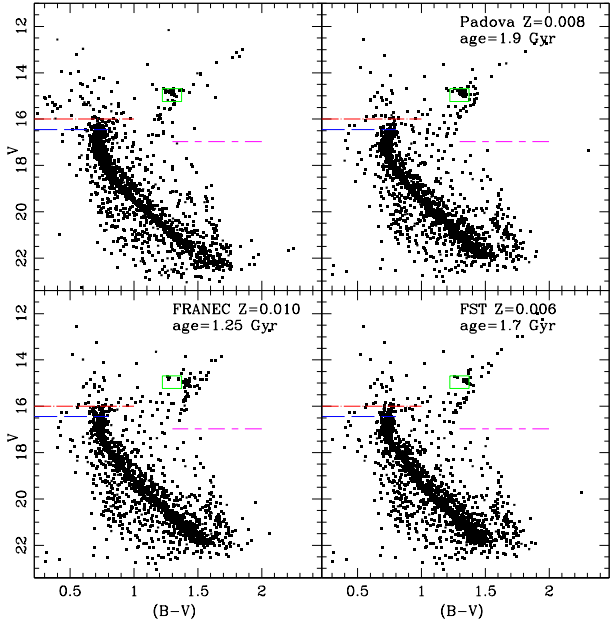}
\caption{Same as Fig. \ref{fig:n1817synth} but for NGC~2141. The observational CMD in the top left panel is for stars inside a 4$\arcmin$ radius area.}
\label{fig:n2141synth}
\end{figure}

Fig.~\ref{fig:n2141synth} shows the comparison between the observed CMD (top left) and the best fits obtained with the three sets of tracks. From this analysis the Padova models provide a better match of the MS shape and of the colour and magnitude of the age indicators. The results are summarised in Tab. \ref{tab:summary}.

\begin{figure}
\includegraphics[scale=0.45]{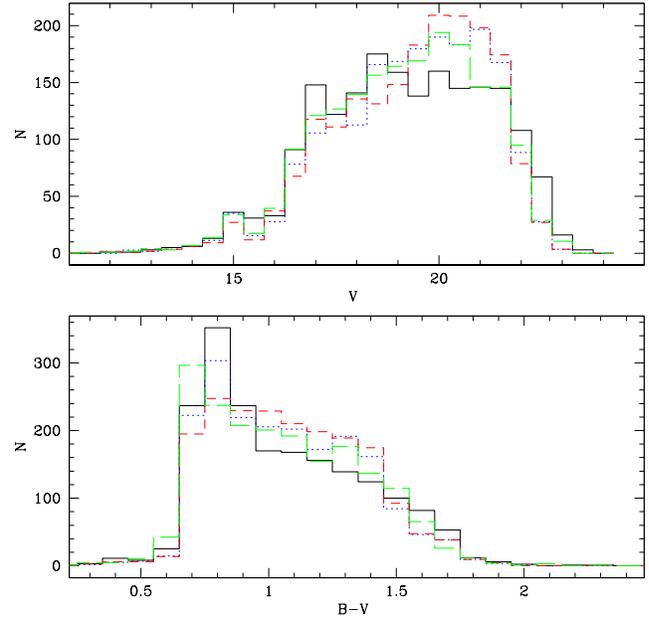}
\caption{Same as Fig. \ref{fig:n1817synth} but for NGC~2141.}
\label{fig:n2141lfs}
\end{figure}

Looking at the luminosity functions of the observational and synthetic CMDs (see Fig. \ref{fig:n2141lfs}) we clearly see that the peak of the 
synthetic distribution is fainter than the observational
one. In the comparison field (shown in Fig. \ref{fig:n2141cmd}) there are clearly MS stars around $V\sim20$. This may be due to 
evaporation. i.e., the typical tendency of low mass stars of moving out of the cluster.
Another possible explanation is related to the Initial Mass Function (IMF).  The best models  predicts a mass of about 0.8 $M_{\odot}$ at $V\sim20$, in the mass range where 
 Salpeter's IMF \citep{salp55} overestimates the mass fraction. Since the synthetic CMDs are generated assuming Salpeter's IMF, they are likely to overpredict  low mass stars. 

Comparing with literature results we find a lower age with respect to both \cite{rosvick95} and \cite{carraro01}. 
In both cases the authors chose a TO
 fainter than ours by about 0.5 mag 
(at about the same level of our RH), and a RC slightly brighter than ours (see Fig. 5 in \citealt{rosvick95}).  Since the age is 
primarily constrained by the magnitude difference between the RC and the MSTO, the large difference in age is explained by the choice of these two age indicators.  We confirm a sub-solar metallicity as suggested by the two papers.

\subsection{Be~81}
\begin{figure}
\includegraphics[scale=0.45]{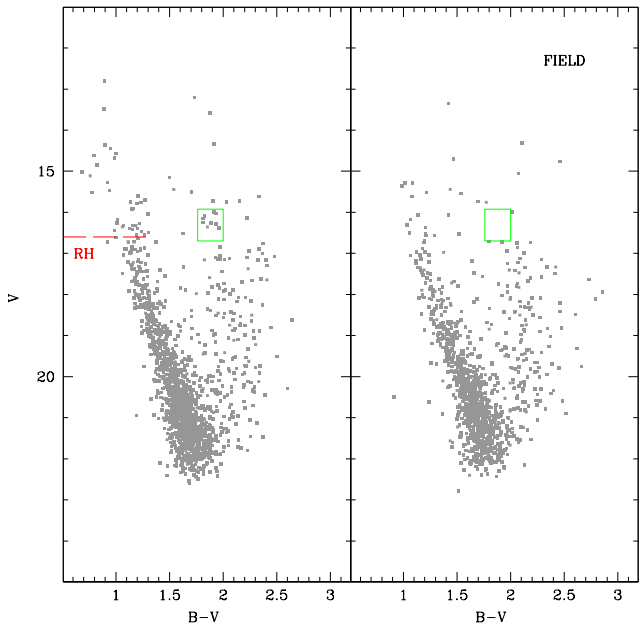}
\caption{{Left panel:} $V,B-V$ CMD for the inner part of Be~81 (inside 2$\arcmin$). The age indicators RC (green box) and RH (red line) are shown. \textit{Right panel:} CMD of the comparison field of the same area. No RC stars appear in the external part of the field.}
\label{fig:be81cmd}
\end{figure}

Be~81 is highly contaminated by field interlopers, condition that makes the interpretation of the cluster features more difficult. For a more robust analysis we studied the inner part of the cluster, where the contrast density with respect to the background density (see Sec. \ref{sec:centre}) is higher and the cluster members should be more evident. From Fig. \ref{fig:be81cmd} an excess at the brighter MS end ($V\simeq15.6$) and on the probable RC locus ($V\simeq16.3$, $B-V\simeq1.8$) is visible for the central part with respect to an external control region. These features have been evaluated for different inner regions and for different choices of comparison field of the same area. We are confident in adopting these features as age sensitive indicators. 

The binary sequence for this cluster is not evident at all from the CMDs because of the high contamination and possibly DR, and for the simulations we adopted a conservative value of 25\% as found on average in many OCs.

We expect a not negligible DR. The MS appears more extended in colour than expected from the photometric error and the probable RC stars have scattered colour and magnitude. After 
several tests,
we decided to adopt a DR of 0.15 for the simulations, with a sensitivity of 0.03. Lower or higher values imply a too tight or too extended MS in the synthetic CMDs.

\begin{figure}
\includegraphics[scale=0.45]{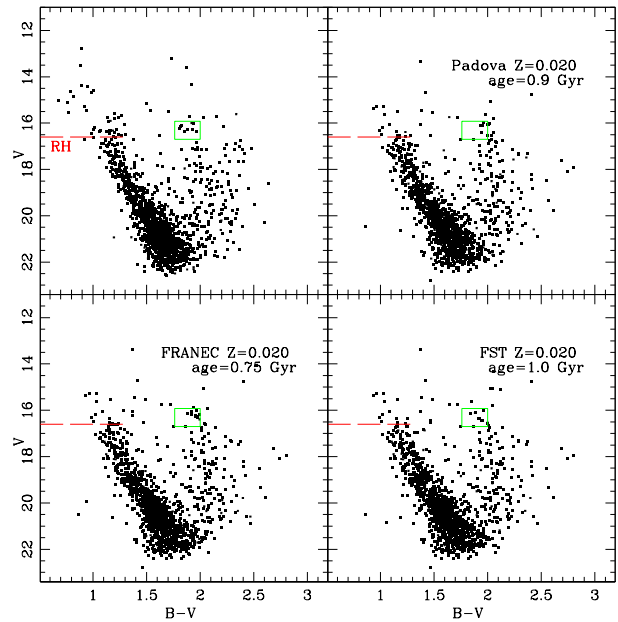}
\caption{Same as Fig. \ref{fig:n1817synth} but for Be~81. The observations CMD in the top left panel is of stars inside a 2$\arcmin$ radius area, corresponding to the half mass radius of the cluster.}
\label{fig:be81synth}
\end{figure}

We find that the cluster footprints (MS and RC) can be reproduced by a solar metallicity, for which we obtain a good match in both $V,B-V$ and $V,V-I$ CMDs. With all the models we can reproduce the magnitude of the age sensitive indicators (RH and RC) and the overall shape of the observational CMD (see Fig. \ref{fig:be81synth}). The colour of the RC is well recovered by FST and FRANEC models. Because of the high contamination from field stars and the effect of severe DR we cannot detail our analysis further.

With the Padova models we find an age of 0.9 Gyr, an average reddening $E(B-V)=0.91$, and a distance modulus $(m-M)_0=12.4$.
In the case of FST models the best match is for an age of 1.0 Gyr, $E(B-V)=0.90$, and $(m-M)_0=12.37$.
With FRANEC we estimate an age of 0.75 Gyr, $E(B-V)=0.92$, and $(m-M)_0=12.45$.

\begin{figure}
\includegraphics[scale=0.45]{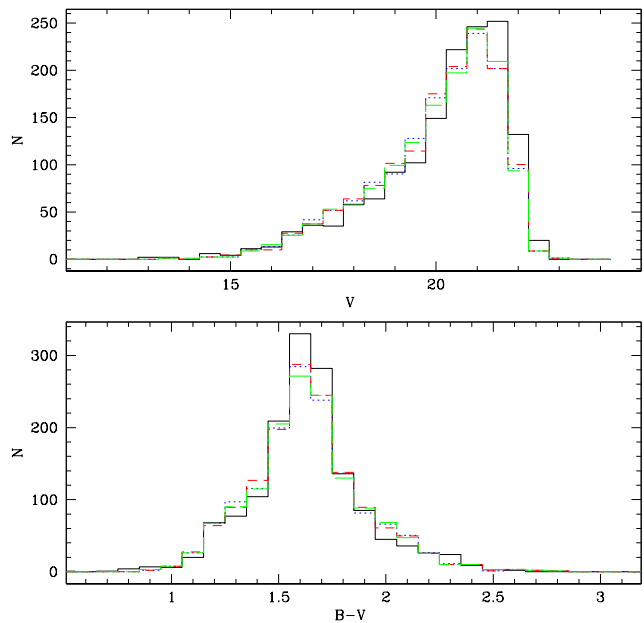}
\caption{Same as Fig. \ref{fig:n1817synth} but for Be~81.}
\label{fig:be81lfs}
\end{figure}

The comparison of the observational and synthetic LFs is very good both in magnitude and in colour, as shown in Fig. \ref{fig:be81lfs}.

We find a good agreement with the results 
presented
 by \cite{sagar98}. We estimate a lower average differential reddening (about 0.1 mag lower) but this can be explained by the differences in our photometries (see Sec. \ref{sec:calib}). On the other hand they exclude that DR 
 be the explanation of the observed broad MS, pointing out that the severe contamination of field stars pollutes the cluster sequences and drives
  the CMD appearance. We investigated further this hypothesis using the synthetic CMD technique and choosing different external areas inside our FoV. We find that the lower MS ($V>19$) is always dominated by field contamination and the signature of the cluster is not evident. 
  Hence we evaluated the DR effect from
   the brighter part of the MS. It is true that there are many field interlopers even for $V<19$, but low DRs always imply
    a too tight synthetic MS and RC with respect to the observations. Hence we  suggest that DR is not negligible across the FoV of Be~81. Firmer conclusions, especially on the cluster metallicity, will be obtained 
    from the analysis of the GES spectra. Both radial velocity measurements and chemical abundance estimates will be fundamental to distinguish cluster members from field stars, cleaning the cluster sequences by interlopers.     

\subsection{The cluster masses}
The synthetic CMD technique can also be used to evaluate the total mass of the clusters summing the masses of all the synthetically generated stars still alive. In order to do that properly we normalised the synthetic population to the star counts inside one $r_h$ and with magnitude $V$ for which 100\% completeness is achieved. The contamination of field stars is taken into account for the normalisation. The derived mass inside one $r_h$ is then multiplied by two to have an estimate of the total mass. The results quoted in Tab. \ref{tab:summary} are obtained with MonteCarlo experiments. We generated 300 hundred synthetics for each cluster taking into account the uncertainty on the normalisation star counts and the error on the distance modulus and differential reddening parameters. The first one is considered as a poissonian error on the counts, hence it affects the number of stars extracted to populate the synthetic. The errors on the distance modulus and reddening (quoted in Sec. \ref{sec:CMDsynth}) affect the mass limit at which the synthetic population is normalised. We use the median of the distribution obtained and its rms. as the reference estimate for the total mass of the cluster.

We can perform this evaluation only for NGC~2141 and Be~81, the two clusters for which we could estimate the King profile (see Sec. \ref{sec:centre}). For NGC~2141 we adopted $V<16.75$ as magnitude limit to normalise the synthetic population. This limit corresponds to the faintest magnitude at which completeness is still 100\% (see Tab. \ref{tab:compl}). For Be~81 we adopted $V<17.25$ as magnitude limit. Using brighter magnitude limits comparable mass estimates are found within the errors.

These computations provide about $1000 M_{\odot}$ for Be~81 and $\sim4000 M_{\odot}$ for NGC~2141.
These mass estimates are a lower limit to the total cluster mass. In fact the stellar models we are using to make synthetic populations have a lower mass limit of 0.6 $M_{\odot}$, hence all the 
 stars with lower mass
  are not taken into account. To get the actual cluster mass we then need to extrapolate along the IMF down to  
0.1 $M_{\odot}$. This implies multiplying by a factor of two the mass if we adopt Salpeter's IMF (Salpeter 1955) and by a factor of 1.4 if we adopt Kroupa's (2002). Since the latter is supposed to best describe the real IMF, we conclude that Be~81 has a mass of $1400 M_{\odot}$ and NGC~2141 of $5600 M_{\odot}$. These are the values listed in Tab. \ref{tab:summary}.

\begin{table*}
  \centering
  \caption{Cluster parameters derived using different models.}
  \begin{tabular}{|l|c|c|c|c|c|c|c|c|c|}
    \hline
    \hline
 Model & age & $Z$ & $(m-M)_0$ & $E(B-V)$ & $d_{\odot}$ & $R_{GC}^a$ &Z & $M_{TO}$ & $M_{tot}$ \\
       & (Gyr) &   & (mag) & (mag) & (kpc) & (kpc)  & (pc) & ($M_{\odot}$) & ($M_{\odot}$)\\
    \hline
    \multicolumn{9}{c}{NGC~1817}\\
    \hline
 Padova & 1.1  & 0.008 & 11.10 & 0.23 & 1.66 & 9.61 & -373.2 & 1.8 & - \\
 FST    & 1.05 & 0.010 & 10.98 & 0.21 & 1.57 & 9.53 & -353.2 & 1.9 & - \\
 FRANEC & 0.80 & 0.010 & 10.88 & 0.34 & 1.50 & 9.46 & -337.3 & 2.0 & - \\ 
    \hline
    \multicolumn{9}{c}{NGC~2141}\\
    \hline
 Padova & 1.9  & 0.008 & 13.20 & 0.36 & 4.37 & 12.21 & -440.9 & 1.5 & $5600\pm300$ \\
 FST    & 1.7  & 0.006 & 13.06 & 0.45 & 4.09 & 11.95 & -413.4 & 1.6 & $6160\pm400$ \\ 
 FRANEC & 1.25 & 0.010 & 13.19 & 0.45 & 4.34 & 12.19 & -438.9 & 1.7 & $4480\pm300$ \\
    \hline
    \multicolumn{9}{c}{Be~81}\\
    \hline
 Padova & 0.9  & 0.020 & 12.40 & 0.91 & 3.02 & 5.74 & -131.3 & 2.1 & $1540\pm100$ \\
 FST    & 1.0  & 0.020 & 12.37 & 0.90 & 2.98 & 5.77 & -129.5 & 2.1 & $1624\pm100$ \\
 FRANEC & 0.75 & 0.020 & 12.45 & 0.92 & 3.09 & 5.69 & -134.4 & 2.2 & $1232\pm100$ \\
    \hline
  \multicolumn{9}{l}{$^aR_{\odot}=8$ kpc is used to compute $R_{GC}$}
  \end{tabular}
  \label{tab:summary}
\end{table*}

\section{Summary and conclusions}\label{sec:sum}
Set in the framework of the BOCCE project \citep[see][]{boc_06}, this paper adds three old open clusters to the BOCCE database. One, 
Be~81,
is located toward the Galactic centre, while the other two, NGC~1817 and NGC~2141, are in the anti-centre direction. They were observed with LBC@LBT using the $BVI$ filters. We obtained CMDs two/three magnitudes deeper than the ones found in literature; hence we could obtain more precise data for the lower MS. The large instrument FoV allowed us to estimate the structure parameters of the clusters NGC~2141 and Be~81 by fitting a King model to their density profile. The analysis of the cluster parameters was carried 
out
using the synthetic CMDs technique that allowed us to infer a confidence interval for age, metallicity, binary fraction, reddening, and distance for each clusters. We used three different sets of stellar tracks (Padova, FST, FRANEC) to describe the evolutionary status of the clusters in order to minimise the model dependence of our analysis. For NGC~2141 a dedicated analysis of the DR is described, using a different technique with respect to the synthetic one and a map of the DR across the cluster is provided. By using the best synthetic CMD and the King profile we evaluated the total cluster mass for NGC~2141 and Be~81.
We found that:
\begin{itemize}
  \item NGC~1817 is located at about 1.6 kpc from the Sun. Its position in the Galactic disc is at $R_{GC}\sim9.6$ kpc and 360 pc below the plane (assuming $R_\odot=8$ kpc as in our previous works). The resulting age 
  is
    between 0.8 and 1.1 Gyr, depending on the adopted stellar model, with better fits for ages between 1.05 and 1.1 Gyr. A metallicity lower than solar seems preferable, in the range $0.006<Z<0.010$. The mean Galactic reddening $E(B-V)$ is between 0.21 and 0.34 and we estimate a 
    fraction of binaries of at least 25\%.
  \item NGC~2141 is at $\sim4.2$ kpc from the Sun, about 12 kpc from the Galactic centre and $\sim$ 430 pc below the Galactic plane. The age is between 1.25 and 1.9 Gyr, with better fits in the age range 1.7-1.9 Gyr. The metallicity for this cluster is lower than solar but higher than $Z=0.004$; the mean Galactic reddening $E(B-V)$ is about 0.40. The estimated binary fraction for this cluster is $\sim$ 16\%. For this cluster we evaluated the effect of the differential reddening: its evident structured MSTO phase, resembling a ``golf club'' shape common to other MW OCs, and its elongated RC can be explained by the presence of not negligible DR across the cluster. The total mass for NGC~2141 is about $5900\pm300M_{\odot}$. 
  \item Be~81 is located toward the Galactic centre at $\sim$3 kpc from the Sun,   
  and at about 130 pc below the plane. Its Galactocentric
  distance $R_{GC}$ is 5.7 kpc . This cluster shows a strong contamination by field stars and an extended MS and RC likely due to differential reddening (up to $0.15$), adding uncertainty to the interpretation of the cluster parameters. The best fitting age is between 0.75 and 1.0 Gyr with a preference for models with a solar metallicity. The reddening estimate is  $E(B-V)\sim0.9$. The total mass of this cluster is $\sim 1500\pm100 M_{\odot}$.
\end{itemize} 
A robust determination of the three clusters parameters would require additional information on cluster membership for evolved and MSTO stars. This is obtainable in the immediate future measuring radial velocities of at least many tens of stars, as in the case of Be~81 within the GES, or we can wait for the results of the Gaia astrometric satellite, with precise individual distances and proper motions. The estimated metallicity is in concordance with their position on the Galactic disc (lower than solar for the outer disc, and solar for the inner part) but only high-resolution spectroscopy will be able to definitely determine the metallicity value.

Our future plan is to update the study described in \cite{boc_06}, adding 
all new BOCCE clusters (we count now 34 OCs), taking into account the information from our studies, the literature, and the on-going surveys, e.g., on metallicity. We will
discuss our findings also in the light of improved models of chemical evolution of the disc and taking into account the latest results on stars and clusters
migration in the disc.

\section*{Acknowledgements}
We thank Paolo Montegriffo, whose software for catalogue matching we
consistently use for our work. For this paper we used  the VizieR catalogue
access tool (CDS, Strasbourg, France), WEBDA, and NASA's Astrophysics Data
System. We are grateful to the LBC team for the pre-reduction procedures.
PD thanks the hospitality of the European Southern Observatory where part of this work was done and the
Marco Polo funds of the Universit\`a di Bologna.

\bsp

\bsp

\label{lastpage}

\end{document}